\documentclass{article}
\usepackage{graphicx}
\usepackage[centertags]{amsmath}
\usepackage{amsfonts}
\usepackage{amssymb}
\usepackage{amsthm}

\title{Space-time as a structured relativistic continuum}

\author{J.~J.~S\l awianowski, V.~Kovalchuk,\\ 
B.~Go\l ubowska, A.~Martens, E.~E.~Ro\.{z}ko\\
Institute of Fundamental Technological Research,\\
Polish Academy of Sciences,\\
$5^{\rm B}$, Pawi\'{n}skiego str., 02-106 Warsaw, Poland\\
e-mails: jslawian@ippt.pan.pl, vkoval@ippt.pan.pl,\\ 
bgolub@ippt.pan.pl, amartens@ippt.pan.pl, erozko@ippt.pan.pl}

\begin{document}

\maketitle

\begin{abstract}
It is well known that there are various models of gravitation: the metrical Hilbert-Einstein theory, a wide class of intrinsically Lorentz-invariant tetrad theories (of course, generally-covariant in the space-time sense), and many gauge models based on various internal symmetry groups (Lorentz, Poincare, ${\rm GL}(n,\mathbb{R})$, ${\rm SU}(2,2)$, ${\rm GL}(4,\mathbb{C})$, and so on). One believes usually in gauge models and we also do it. Nevertheless, it is an interesting idea to develop the class of ${\rm GL}(4,\mathbb{R})$-invariant (or rather ${\rm GL}(n,\mathbb{R})$-invariant) tetrad ($n$-leg) generally covariant models. This is done below and motivated by our idea of bringing back to life the Thales of Miletus idea of affine symmetry. 
Formally, the obtained scheme is a generally-covariant tetrad ($n$-leg) model, but it turns out that generally-covariant and intrinsically affinely-invariant models must have a kind of non-accidental Born-Infeld-like structure. Let us also mention that they, being based on tetrads ($n$-legs), have many features common with continuous defect theories. It is interesting that they possess some group-theoretical solutions and more general spherically-symmetric solutions. It is also interesting that within such framework the normal-hyperbolic signature of the space-time metric is not introduced by hand, but appears as a kind of solution, rather integration constants, of differential equations. Let us mention that our Born-Infeld scheme is more general than alternative tetrad models. It may be also used within more general schemes, including also the gauge ones.
\end{abstract}

\section*{Introduction}

In this paper we deal with first-order variational principles for the field of linear co-frames in an $n$-dimensional manifold. We formulate a certain class of models with Lagrangians invariant both under the group of diffeomorphisms and under the matrix group ${\rm GL}(n,\mathbb{R})$ acting in a natural way on linear frames. This internal ${\rm GL}(n,\mathbb{R})$-symmetry is the main difference between models presented here and relativistic theories of micromorphic continua or metric-teleparallel theories of gravitation (including the conventional Einstein theory). The latter theories are invariant only under the Lorentz subgroup ${\rm SO}(1,n-1;\mathbb{R})$ acting globally in the general case and locally in the Einstein theory. We show that from some point of view the restriction of the internal group ${\rm GL}(n,\mathbb{R})$ to ${\rm SO}(1,n-1;\mathbb{R})$ is artificial.

We give some heuristic arguments which seem to support the hypothesis that perhaps the presented formalism is a promising way toward a geometric theory of fundamental interactions (a new description of gravitation or some kind of a unified field theory). Our variational principles can be easily generalized to systems of $m$ covector fields with $m>n$, i.e., to covector fields with higher-dimensional internal isotopic space.

We present simple examples of rigorous solutions of our field equations. There exists a link between these solutions and group-theoretical structures in manifolds. Every field of linear frames whose ``legs'' span a semisimple Lie algebra is a solution. The manifold then becomes a homogeneous space of a freely acting semisimple Lie groups. Such solutions play in our model the same role as the flat-space solutions in the Einstein theory and in metric-teleparallel theories of gravitation. Roughly speaking, they are ``homogeneous vacuums'', i.e., classical ground states of our model. Trivial central extensions of semisimple Lie groups give rise to another class of natural  solutions (``homogeneous-developing vacuums'').

Preliminary discussion of isotropic solutions is also presented. We formulate the general procedure of the search for isotropic solutions with a factorized dependence on variables $t$, $r$. The time-dependence is exponential and the resulting pseudo-Riemannian structure is stationary, although non-static. It is interesting that the normal-hyperbolic signature is not a priori assumed (introduced ``by hand''). It is a property of the most natural solutions of our field equations. Roughly speaking, the signature is implied by differential equations. There are certain similarities between models presented here and nonlinear models of electrodynamics due to Born, Infeld and Mie, cf. also \cite{Chr1,Chr2,Chr3,Chr4,Chr5,PG_02,PG_03_1,PG_05,PG_08,PG_10,PG_13,11,12,JJS_10,all_10,all_12_2}.

One thing must be stressed here. During a few last decades we witnessed a triumphal progress of gauge theories in physics. The best example is the standard model of electroweak interactions and also the chromodynamical theory of strong interactions. The gauge idea was also successfully applied in the theory of condensed matter, first of all in superconductivity, in superfluids theory and also in the theory of defects in elastic continua \cite{Kro_29,AT+JJS_90}. These successes suggested also to do some attempts of the gauge formulation of gravitation theory, cf., e.g., \cite{5-1,5-2,Ivan_146,10,Ivan_173,Ivan_35,Ivan_124,Ivan_123,PBO_85,25}. The idea looks natural from the point of view of the unity of physics, the more so, the standard Einstein general relativity, in spite of its success and unquestionable value, has some weak points as well, first of all the notorious non-renormalizability on the quantized level. There are various models of gauge theories of relativity using various gauge groups: the internal Lorentz or rather Poincare group, the internal conformal group or rather its covering ${\rm SU}(2,2)\in {\rm GL}(4,\mathbb{C})$, the linear group ${\rm GL}(4,\mathbb{R})$, the affine group ${\rm GAff}(4,\mathbb{R})$ and even the complex group ${\rm GL}(4,\mathbb{C})$. The corresponding theories are formulated in appropriate bundles over the space-time manifold, e.g., in principal bundles of Lorentz-orthonormal frames, general linear frames, etc. And the resulting schemes are invariant under the local, i.e., $x$-dependent action of the gauge groups. Of course, those symmetry groups are infinite-dimensional due to the dependence on the space-time point $x$. The resulting theories are essentially nonlinear and in general very complicated. However, as stressed by some authors, some kind of approximation may be obtained when one restricts ourselves to globally-invariant, i.e., $x$-independent schemes. For example, it is so for all versions of the tetrad models of gravitation. Their predictions are qualitatively compatible with those of gauge theories and of general relativity. The only feature to be preserved is their general covariance, i.e., invariance under the total group of space-time diffeomorphisms.

But there are also some other reasons for such models. First of all, one can think about the dynamics of the field of frames in a higher-dimensional manifold. And then the local invariance in the sense of the four-dimensional space-time might be in a sense derived as a consequence of the global ${\rm GL}(n,\mathbb{R})$-invariance in the higher-dimensional ``Kaluza-Klein space-time''. But there is also another, perhaps more important, motivation. It has to do with the big-bang philosophy. Namely, according to the big-bang scenario, in the first moments of the Universe evolution, the global ${\rm GL}(4,\mathbb{R})$-symmetry may be more essential than the local, i.e., $x$-dependent one. But then the globally ${\rm GL}(4,\mathbb{R})$-invariant model may have a chance to be physically more justified (the more so, the ${\rm GL}(n,\mathbb{R})$-globally invariant models with $n>4$).

Finally, let us finish this introduction section with the remark that the tetrad ($n$-leg) models are also interesting from the point of view of analogy between field theory and continuum mechanics, including defects. They provide also a good illustration to the usefulness of the general method of non-holonomic frames \cite{Sch_54,JJS_91,JJS_29,JJS_30,JJS_98,JJS+VK_02,VK_04}.

\section{Hypothetical carrier of fundamental interactions: field of frames}

Let $M$ be a smooth and orientable $n$-dimensional ``space-time'' manifold. Mathematical universe of the classical theory of bosonic fields consists of the principal fiber bundle of linear frames $FM$ and of its associated vector bundles. Physical fields are represented by cross-sections of the appropriately chosen fibered structures. Of particular interest are the tensor bundles over $M$.
Let us consider a field whose kinematics is given by a fiber bundle $(E,M,\pi)$,
$\pi:E\rightarrow M$ being the projection. Realistic theories are based on Lagrangians of the first differential order, thus, the proper mathematical framework for the dynamics is the manifold $J(\pi)$ of first-order jets of $E$ over $M$ (we do not write $J^{1}(\pi)$ because no higher-order jets will be used). $J(\pi)$ is a bundle both over $E$ and $M$ with the natural projections
$\tau: J(\pi)\rightarrow E$, $\theta=\pi\circ\tau: J(\pi)\rightarrow M$. 
Any field, i.e., any cross-section $\sigma: M\rightarrow E$, may be lifted in a natural way to $J(\pi)$, resulting in a cross-section $j\sigma: M \rightarrow J(\pi)$.

From the geometric point of view Lagrangian is a $\theta$-vertical differential $n$-form $\lambda$ on $J(\pi)$. Any cross-section $\sigma: M\rightarrow E$ associates with $\lambda$ a differential $n$-form $\mathcal{L}[\sigma]$ on the base manifold $M$, $\mathcal{L}[\sigma]:=(j\sigma)^{\ast}\cdot\lambda$.
Conversely, $\sigma$ and $\mathcal{L}[\sigma]$ determine $\lambda$ in a unique way. We say that $\mathcal{L}[\sigma]$ is a Lagrangian along the field evolution $\sigma$.
If $\Omega\subset M$ is a regular $n$-dimensional region, then the action over $\Omega$ along the field evolution $\sigma$ is given by
\begin{equation}
I\left[\Omega,\sigma\right]:=\int_{\Omega}\mathcal{L}[\sigma]=
\int_{(j\sigma)(\Omega)}\lambda.
\end{equation}
In general $\mathcal{L}$ involves both the dynamical variables $Y$ and the absolute objects $G$, thus, $\sigma=\left(G,Y\right)$ and we shall use the symbols $I\left[\Omega,G,Y\right]$, $\mathcal{L}\left[G,Y\right]$.

In the Hamilton principle of stationary action the quantities $Y$ are subject to the variation procedure; on the contrary, the objects $G$ are kept fixed. The field evolution $Y$ can actually occur in Nature iff for any region $\Omega$ it gives a stationary value to the functional $I\left[\Omega,G,\cdot\right]$ within the class of all virtual fields $\psi$ which coincide with $Y$ on the boundary $\partial\Omega$, $\left.\psi\right|_{\partial\Omega}=\left.Y\right|_{\partial\Omega}$.
This results in the familiar Euler-Lagrange equations to be satisfied by $Y$. The particular shape of $G$ is a ``parameter'' of the dynamics for $Y$. If $\mathcal{L}$ involves absolute quantities, then the physical system described by $Y$ is dynamically open. The family of solutions is then controlled by $G$.

Let $\phi: M\rightarrow M$ be a diffeomorphism of $M$ onto itself. If the theory is correctly formulated, i.e., if there are no hidden absolute objects, then we have
$\mathcal{L}\left[\phi^{\ast}G,\phi^{\ast}Y\right]=
\phi^{\ast}\mathcal{L}[G,Y]+d\Gamma[G,Y]$,
where $\Gamma$ is a differential $(n-l)$-form on $M$ built algebraically of $G$ and $Y$, while $\phi^{\ast}$ denotes the $\phi$-transformation of fields meant in the pull-back convention.
In other words we have
$I\left[\phi\Omega;\phi_{\ast}G,\phi_{\ast}Y\right]=I\left[\Omega;G,Y\right]
+\Delta\left[G|_{\partial\Omega},Y|_{\partial\Omega}\right]$,
where $\Delta$ is a functional defined on the family of boundary values of $G$ and $Y$ on $\partial\Omega$, while $\phi_{\ast}$ is a push-forward transformation of tensor fields.
Therefore, if $Y$ satisfies the field equations with the absolute object $G$, then $\phi_{\ast}Y$ satisfies the field equations with the absolute object $\phi_{\ast}G$. In general, $\phi_{\ast}Y$ does not fulfil the original field equations based on $G$. Thus, $\phi\in{\rm Diff}(M)$ is a dynamical symmetry iff $\phi^{\ast}G=G$, i.e., if it preserves the absolute quantities.

The absolute objects $G$ have a double physical meaning: macroscopic and microscopic. On the phenomenological macroscopic level they describe the directly observed geometry of the physical space-time (distances, angles, time intervals, parallel transport, volume, etc.). The microscopic interpretation of $G$ is simply given by its position in the action functional. Physical fields included into account of degrees of freedom are brick-stones of the Lagrangian; on the other hand the absolute objects are used as its ``skeleton''. They appear as auxiliary quantities necessary to ``glue'' dynamical variables and their derivatives into scalars and scalar densities used in the action functional. Obviously, the macroscopic meaning of absolute objects is a consequence of their ``microscopic'' position in variational principles.

The particular choice of absolute objects included into theory depends on the nature of considered problems. Typical realistic theories are based on the following triple of absolute quantities: $(i)$ the metric tensor $g$, $(ii)$ the affine connection $\Gamma$, $(iii)$ the standard of volume and orientation represented by a nowhere vanishing differential $n$-form $\varepsilon$. Usually they are not independent on each other. The most economic and most popular model is that based on $g$ alone; $\Gamma$ and $\varepsilon$ are respectively the natural Levi-Civita connection and the natural pseudo-Riemannian volume element (orientation itself must be fixed independently).

The task of $g$ is to shift and to contract the tensor indices; these operations are necessary for obtaining scalars from tensors. Affine connection enables us to differentiate tensor fields. The oriented standard of volume occurs in integral formulas, first of all in the action functional, 
$I_{\Omega}[F]=\int_{\Omega}\Lambda\left(F,\nabla F\right)\varepsilon$, 
where the system of tensors $F$ describes dynamical variables of the theory and $\Lambda$ is a scalar field built algebraically (with the help of $g$) of $F$ and of its covariant derivative $\nabla F$. The quantity $\Lambda$ is responsible for the dynamical structure of the theory. It describes the density of action with respect to the volume standard $\varepsilon$.

Let us stress the following important point: In theories with the absolute geometry of the type $\left(g,\Gamma,\varepsilon\right)$ it is always possible to construct a first-order Lagrangian for any kind of tensorial field. The most economic model is that with $\Gamma$ and $\varepsilon$ built of $g$ (Levi-Civita connection and pseudo-Riemannian volume). Thus, the most fundamental
and universal absolute object is the metric tensor alone.
According to the Einstein-Hilbert postulate of the general covariance, we believe that the actually fundamental theories should not involve absolute objects at all. Any quantity occurring in the Lagrangian $\mathcal{L}$ belongs to physical degrees of freedom and should be subject to the variation procedure when we derive the Euler-Lagrange field equations. Fundamental theories are invariant under ${\rm Diff}(M)$, i.e.,
$I\left[\phi\Omega,\phi_{\ast}Y\right]=I\left[\Omega,Y\right]+
\Delta\left[Y|_{\partial\Omega}\right]$,
(cf. \cite{31,32,33}) or, equivalently,
$\mathcal{L}\left[\phi^{\ast}Y\right]=\phi^{\ast}\mathcal{L}[Y]+d\Gamma[Y]$.
If $Y$ is a solution of the Euler-Lagrange equations, then so is $\phi^{\ast}Y$ for any $\phi\in {\rm Diff}(M)$.

When we give up the absolute objects and go over to the generally-covariant framework, then the fields and the associated bundles of $FM$ admitting nontrivial Lagrangians become rather exceptional (try to construct a generally-covariant Lagrangian for the scalar field or for the covector field). It is rather typical that the requirement of the invariance under ${\rm Diff}(M)$ is incompatible with degrees of freedom. Therefore, the Einstein-Hilbert programme leads in a natural way to the following questions: 1.\ What are exceptional associated bundles of $FM$, i.e., exceptional kinds of physical fields, admitting variational principles invariant under ${\rm Diff}(M)$ and involving first-order derivatives? In more physical terms: which kinds of elementary particles can exist autonomously in a bare, structureless, manifold? 2.\ Is it possible to extend those exceptional fields to some larger self-inter\-acting systems within the associated universe of $FM$? 3.\ Is this universe ordered in a hierarchic way? If yes, what are the most fundamental objects? (fundamental particles?) 4.\ Does there exist a universal network of couplings for all tensorial and spinorial fields? When analyzing these problems we could in principle appeal to the general mathematical theory developed by Krupka and others \cite{19,20,18}. However, for our purposes it is more convenient to use simple intuitive arguments and to develop variational models on the independent basis.

The simplest way to seek a generally-covariant theory is based on the ``elastization'' of the ``absolute'' Lagrangian $\mathcal{L}\left[G,F\right]$. Namely, we keep $\mathcal{L}$ unchanged and decide to regard $G$ and $F$ on the same footing, as dynamical variables subject to the variation procedure. As a rule, the theory based on $\mathcal{L}\left[G,F\right]$ would be non-satisfactory, both physically and mathematically. For example, if $G$ is the metric tensor, then the subsystem of the Euler-Lagrange equations resulting from the variation of $G$ in $\mathcal{L}\left[G,F\right]$ reads $T=0$, i.e., the symmetric energy-momentum tensor of $F$ would have to vanish. This is physically incorrect and in general mathematically incompatible with the remaining Euler-Lagrange equations (dynamical equations for $F$). As a Lagrangian for the total system $\left(G,F\right)$, the quantity $\mathcal{L}\left[G,F\right]$ would be strongly singular and the resulting system of the field equations would be over-determined and in general inconsistent. The standard historical way to overcome this difficulty is to interpret the original quantity $\mathcal{L}\left[G,F\right]$ as a ``matter Lagrangian'' $\mathcal{L}_{m}$ of the ``physical'' fields $F$ influenced by the ``gravitational'' (geometrical) quantities $G$. The total Lagrangian of the system $\left(G,F\right)$ is expected to have the form $\mathcal{L}\left[G,F\right]=\mathcal{L}_{m}\left[G,F\right]+
\mathcal{L}_{g}\left[G\right]$,
where $\mathcal{L}_{g}$ is a first-order Lagrangian for ``gravitational'' quantities. This is just the minimal-coupling scheme for $\left(G,F\right)$. It turns out that realistic (practically used) $G$-s admit first-order variational principles invariant under ${\rm Diff}(M)$. For example, if $G$ consists only of the metric $g$, then $\mathbb{L}_{g}=R\sqrt{g}$, i.e., locally, in a chart $\left(U,x^{1},\ldots,x^{n}\right)$,
$\left.\mathcal{L}_{g}[g]\right|_{U}=R\sqrt{|\det\|g_{ij}\||}dx^{1}\wedge\ldots\wedge dx^{n}$, where $R$ is the curvature scalar of the pseudo-Riemannian manifold $\left(M,g\right)$. Up to the ``cosmological'' correction proportional to $\sqrt{g}$, this Einstein-Hilbert Lagrangian is the only possibility. $\mathcal{L}_{g}$ is essentially a first-order Lagrangian because the second-order terms contained in $R$ form a total-divergence expression. There are also Palatini-like models (affine-metric theories of gravitation among them \cite{6-1,6-2,6-3,7,8,9}) where the gravitational degrees of freedom are described by the pair $\left(g,\Gamma\right)$, i.e., by the metric tensor and affine connection considered as independent variables. There are also minimal models opposite to the Einstein-Hilbert model, namely those based on the affine connection alone as a gravitation variable. Such theories were investigated by Eddington, Schr\"odinger and Kijowski \cite{15}. All those models admit first-order Lagrangians invariant under ${\rm Diff}(M)$.

No doubt the dominant theory is the conventional Einstein relativity based on $g$ alone as a ``gravitational'' variable. It is simplest in that it does not involve ``metafields'' like the connection form which is not an inhabitant of the universe of usual bundles over $M$ (the connection form is an $L\left(n,\mathbb{R}\right)$-valued differential form on $FM$). The most popular answer given by physicists to the above questions 1--4 is based on the Einstein theory and it may be formulated as follows. The metric tensor $g$ is a minimal self-interacting field. The assumption of a first-order variational dynamics distinguishes the bundles of symmetric second-order tensors $ST^{0}_{2}M$, $ST^{2}_{0}M$ as dominant elements of the universe of all associated bundles of $FM$. Gravitons are the most fundamental bosons. They exist autonomously in a bare manifold. Any system of fields including $g$ admits a ${\rm Diff}(M)$-invariant variational dynamics, i.e., it can exist without any absolute objects. Lagrangians of all ``physical'' fields involve $g$, thus, the metric tensor provides a mathematical description of the ``universal gravity'' through which all kinds of fields (elementary particles) are coupled together even if all other interactions are ``switched off''. The fundamental character of gravity is reflected exactly by the fact that it can be never ``switched off''. $\mathcal{L}_{g}$ is not quadratic in $g$, thus, the field $g$ is a self-interacting (self-gravitating) kernel of the physical reality.

There exists a popular opinion that there are no other kernels of this type. In this way the bundles $ST^{0}_{2}M$, $ST^{2}_{0}M$ acquire the physically privileged position within the bosonic universe. We are faced with a characteristic dualism of two kinds of fields: geometrical (gravitational) and physical (material). The first group consists of the metric tensor and of its concomitants. This means that the metric tensor $g$ seems to be something fundamental and exceptional, exactly as in theories involving absolute objects. It survives the Einstein-Hilbert revolution (the postulate of general covariance).

One can object against thin ``metrical creed'' for both geometrical and physical reasons. First of all, the bundles $T^{0}_{2}M$, $T^{2}_{0}M$ are rather accidental and certainly not very fundamental members of the bundle universe of $FM$ (although they are low-valence objects and have a natural interpretation in terms of fundamental bundles $TM$, $T^{\ast}M$; namely, they describe homomorphisms between vectors and covectors). If the set of all possible bosonic
fields is to have a geometric ``kernel'', then certainly the best candidate is its natural geometric ``ruler'', i.e., the principal bundle $FM\simeq F^{\ast}M$. It is reasonable to suspect that the dominant role of the metric tensor (and of the affine connection) in generally-covariant field theories is a consequence of our habit of using the absolute geometry. Einstein theory provides the simplest procedure (so to speak, the minimal-change procedure) leading from physics with the absolute objects to the ${\rm Diff}(M)$-invariant framework. However, if we decide to give up the absolute quantities, then it seems reasonable to ``forget'' traditional concepts of macroscopic geometry (metric tensor, affine connection, and all that). There is no geometry any longer, there are only elementary particles and their classical fields in the physical space-time.

Therefore, one should review the universe of bosonic fields and try to answer the above questions 1--4 without any prejudices. Of course, a priori $FM\simeq F^{\ast}M$ seems to be the best candidate for the fundamental object because it is a principal fiber bundle and all other bundles of the bosonic universe are its associated bundles. Roughly speaking, the bundle $FM\simeq F^{\ast}M$ absorbs the geometry of this framework. The ``tetrad'' field, i.e.,
the cross-section of $FM$ is a genuine geometric quantity, while the other fields can be always represented by their scalar components with respect to a fixed field of frames. More complicated geometric objects, e.g., tensor densities, are defined through their transformation properties with respect to the tetrad deformation. This dominant and geometrically privileged role of $FM$ is obscured by the fact that the tensor bundles over $M$ are natural with respect to $M$; they are, so to speak, ``soldered'' to the manifold $M$ \cite{34,35}. Moreover, $FM$ is explicitly constructed from its own associated bundle, namely from $TM$. This soldering effect completely disappears when we try to construct half-objects necessary for the theory of fermionic fields. There is no canonical principal bundle over $M$ with the covering group $\overline{{\rm GL}\left(n,\mathbb{R}\right)}$ as a structural group. Similarly within the four-dimensional metrical framework there is no canonical principal bundle over $M$ with ${\rm SL}\left(2,\mathbb{C}\right)$ as a structural group.

Let us quote a few additional and more physical arguments in favour of $FM$:
1.\ We are used to the reductionist methodology which advises us to explain everything in terms of elementary and non-divisible entities. On the basis of this methodology $FM$ seems to be better than $ST^{0}_{2}M$, i.e., the $n$-tuple of vector bosons seems to be more elementary than the non-intuitive spin-two particle, although it has more components (namely, $n^{2}$ instead of $n(n+1)/2$). Indeed, all tensors can be constructed from vectors but not conversely. We expect that the actually fundamental particles should have smallest nontrivial values of spin.
2.\ Gauge theories seem to teach us that the really fundamental interactions should be carried by covector bosons. Incidentally, the idea of a universal and geometric interaction carried by the quadruple of vector bosons seems to be interesting in the light of the Salam-Weinberg model.
3.\ When we introduce spinor fields describing fermionic matter, then the tetrad field is a necessary tool (or, at least, very convenient one). This use of $FM$ is extremely important.
4.\ ${\rm GL}\left(n,\mathbb{R}\right)$ is the structural group of $FM$. At the same time, linear geometry, ruled by the $n$-dimensional linear group, is a fundamental and most elementary geometry in tangent spaces of a differentiable manifold, prior to any extra introduced structure. Thus, it is a tempting idea to regard as a fundamental physical object in $M$ something that ``feels'' the action of ${\rm GL}\left(n,\mathbb{R}\right)$, i.e., the tetrad field. Space-time then becomes a ``micromorphic continuum'', i.e., its elementary constituents are infinitesimal homogeneously deformable grains (tetrads) \cite{JJS_86,JJS_28,JJS_03}.
5.\ There are serious attempts to interpret ${\rm GL}\left(4,\mathbb{R}\right)$ and ${\rm GL}\left(3,\mathbb{R}\right)$ as fundamental symmetries in elementary particles physics \cite{25,8,9}.

On the basis of the above arguments we simply feel to be forced to formulate the following programme:
1.\ The cross-section of $FM$, i.e., the field of linear frames, is a candidate for the fundamental physical field admitting a ${\rm Diff}(M)$-invariant variational dynamics. (Thus, the bundle $FM$ is a candidate for the self-interacting kernel of the physical reality).
2.\ We also need the metric tensor, first of all in algebraic operations of tensorial contraction, when coupling together physical fields into Lagrangians and interaction terms. Metrical concepts are used also in the standard theory of spinors. Thus, the theory we seek should involve some metric tensor, however this tensor will be no longer an autonomous degree of freedom (as it is in the Einstein theory), but instead it should be a secondary quantity constructed of the field of frames.
3.\ We search for a theory invariant under the structural action of ${\rm GL}\left(n,\mathbb{R}\right)$ on $FM$, because this action describes the natural kinematic symmetries of degrees of freedom.

There are two objections which could be a priori raised against the above programme. Let us formulate them and try to answer them before the formal developing of our theory:

\noindent {\bf Objection 1:}  The program is superfluous because in fact there exist field theories which attribute gravitational degrees of freedom to the principal bundle $FM$. As a typical example let us mention the gauge-theoretic formulation with Lagrangians quadratic in curvature. Gravitation quantities are represented by the connection form. Introducing an auxiliary tetrad field we can represent the connection form by a system of $n^{2}$ differential forms on $M$, i.e., covector bosons. Therefore, one could claim that this formulation just satisfies the requirement of the dominant position for $FM$ or for the first-floor tensorial objects like $TM$, $T^{\ast}M$.

\noindent{\bf Answer:} The connection form is not defined on $M$ but on $FM$, thus, it is not a field in the usual sense. We can represent it by a system of covector fields on $M$ only after the tetrad field is introduced. The latter appears then as a merely coordinate-like auxiliary variable, not as a dynamical field. Thus, just as other Palatini-like theories, this formulation does not fulfil our requirements. Moreover, the metric field is still a necessary and unavoidable dynamical variable, because without it the quadratic Lagrangian could not be
constructed. Let us mention however that among all theories of gravitation using affine connection as a primary quantity those formulated by Eddington and Kijowski \cite{15} are free of the metrical degree of freedom. Kijowski uses the Lagrangian $n$-form which is locally given by 
$\sqrt{|\det\|R_{ij}\||}dx^{1}\wedge\ldots\wedge dx^{n}$, 
where $R_{ij}$ is the Ricci tensor of the affine connection. The metric field appears then as a secondary variable through canonical momenta conjugate to the connection form. Among all theories of gravitation this one is nearest to our ideas. \rule{5pt}{5pt}

\noindent{\bf Objection 2:} The programme is superfluous because the familiar tetrad formulation of the conventional Einstein relativity and its metric-tele\-parallel generalizations provide us with the framework in which $FM$ acquires the required dominant position and the tetrad field is a fundamental degree of freedom. The metric tensor then appears as a secondary field algebraically built of the tetrad by the pointwise injection of the numerical Minkowskian metrics,
\begin{equation}\label{1.1}
g=\eta_{AB}\phi^{A}\otimes\phi^{B},\qquad \eta_{AB}={\rm diag}\left(1,-1,\ldots,-1\right).
\end{equation}
This formulation has an additional advantage, namely, we can use Lagrangians represented by scalar densities explicitly free of second derivatives.

\noindent{\bf Answer:} Einstein theory formulated in the tetrad terms is invariant under the local action of the Lorentz group ${\rm SO}\left(1,n-1;\mathbb{R}\right)$ (by ``local'' we mean $M$-dependent). Therefore, that part of the tetrad field which does not contribute to $g$ is a non-physical variable which is not accounted to degrees of freedom and can be given any a priori required form. This means that the tetrad field is not a fundamental physical quantity; the local Lorentz symmetry is too strong. The general metric-teleparallel theories \cite{17,22,23} are free of this disadvantage. However, they do not satisfy our requirements because the pointwise relationship (\ref{1.1}) between $\phi$ and $g$ implies that they are invariant under the global action of ${\rm SO}\left(1,n-1;\mathbb{R}\right)$, but not under the total ${\rm GL}^{+}\left(n,\mathbb{R}\right)$. The local Lorentz symmetry would be too strong, but the global one is too weak. Indeed, degrees of freedom are ruled by ${\rm GL}\left(n,\mathbb{R}\right)$ and it is difficult to see a sufficient reason for the aprioric restriction of the required ${\rm GL}\left(n,\mathbb{R}\right)$-dynamical symmetry to ${\rm SO}\left(1,n-1;\mathbb{R}\right)$. This restriction would be equivalent to the non-motivated introduction of the additional primitive element $\eta$, or
equivalently, to the non-motivated assumption that the metric field should be covariantly constant under the parallelism connection induced by the tetrad field. In any case, there is no link between the reduction of ${\rm GL}^{+}\left(n,\mathbb{R}\right)$ to ${\rm SO}\left(1,n-1;\mathbb{R}\right)$ and the requirement of the hyperbolic signature of the metric tensor $g[\phi]$ built of $\phi$. Such a link is characteristic for the algebraic relationship (\ref{1.1}) between $\phi$ and $g[\phi]$, but there is no internal necessity to put $g[\phi]$ in the form (\ref{1.1}). \rule{5pt}{5pt}

Below we show that there exist mathematically nontrivial ${\rm GL}^{+}\left(n,\mathbb{R}\right)$-inva\-riant variational principles with the field of frames as a fundamental self-inter\-acting and universally-coupling field. Thus, within the universe of bosonic geometric objects on $M$ there exist two main a priori possible universal channels of interaction: the bundle $ST^{0}_{2}M$ of symmetric second-order tensors and the bundle $FM$ of linear frames.

The corresponding geometric objects are two-indices ones, namely the twice covariant tensors and the $n$-tuples of (co-)vectors. Let us notice, however, that they are not the only possible objects admitting generally covariant Lagrangians. There are yet another two kinds of such objects: mixed second-order tensors and fields of $n\times n$ matrices on the $n$-dimensional manifold. They were partly mentioned in \cite{JJS_3}. Let us quote here some additional remarks. First let us repeat the concept of Nijenhuis torsion. With any part of second-order mixed tensors $X$, $Y$ one can associate the third-order skew-symmetric tensor $S(X,Y)$, i.e.,
$S(X,Y)^{\mu}{}_{\nu\lambda}=-S(X,Y)^{\mu}{}_{\lambda\nu}$,
symmetric in tensors $X$, $Y$: $S(X,Y)=S(Y,X)$.
The structure of $S(X,Y)$ enables one to interpret $S(X,Y)$ as a mapping which assigns a vector field $S(X,Y)\cdot(A,B)$ to any pair of vector fields $A$, $B$, i.e., $S(X,Y)\cdot(A,B)=S(X,Y)^{\mu}{}_{\nu\lambda}
A^{\nu}B^{\lambda}\partial/\partial x^{\mu}$. It is defined as
$S(X,Y)\cdot(A,B)=[XA,YB]+[YA,XB]+XY[A,B]+YX[A,B]
-X[A,YB]-X[YA,B]-Y[A,XB]-Y[XA,B]$.
Obviously, for every pair of vector fields $C$, $D$, the symbol $[C,D]$ denotes the Lie bracket, $[C,D]=CD-DC$, i.e., analytically
$[C,D]^{\mu}=C^{\nu}\partial_{\nu}D^{\mu}-D^{\nu}\partial_{\nu}C^{\mu}$.
The composition $XY$ of linear mappings is meant in the pointwise sense, i.e., in terms of coordinates:
$(XY)^{\alpha}{}_{\beta}=X^{\alpha}{}_{\mu}Y^{\mu}{}_{\beta}$.
The definition of $S(X,Y)$ as a $T^{1}_{2}$ tensor is correct because in spite of the apparent dependence of the result on derivatives they mutually cancel with each others.

If we substitute $Y=X$, then we obtain the quantity $S(X):=S(X,X)$.
It is also possible to construct the higher-order quantities like, for instance,
$S^{k,l}(X)=S\left(X^{k},X^{l}\right)=S^{l,k}(X)$.
Obviously, the zeroth-order ones equal $X^{0}={\rm Id}$, $S^{0,1}(X)=S^{1,0}(X)=0$.
The simplest choice is $S(X)=S(X,X)=S^{1,1}(C)$.
It is linear both in $X$ and the system of first derivatives of $X$. Let us mention that the dependence on derivatives is linear. The same concerns the dependence on the field $X$ itself. It is clear that for any diffeomorphism $\varphi$ of the space-time onto itself,
$S\left(\varphi_{\ast}X\right)=\varphi_{\ast}S(X)$.
The best and simplest candidate for the Lagrange tensor is 
$\mathcal{L}(X,\partial X)_{\mu\nu}=AS^{\lambda}{}_{\mu\varkappa}S^{\varkappa}{}_{\nu\lambda}
+BS^{\lambda}{}_{\mu\lambda}S^{\varkappa}{}_{\nu\varkappa}+
CS^{\lambda}{}_{\varkappa\lambda}S^{\varkappa}{}_{\mu\nu}$.
The corresponding Lagrangian will be given by the square root of the determinant of $\mathcal{L}(X,\partial X)_{,\mu\nu}$, i.e.,
$L=\sqrt{|{\rm det}\left[\mathcal{L}(X,\partial X)_{\mu\nu}\right]|}$.
It is geometrically correct (the Weyl tensor density of weight one, obtained by the square-rooting of the second-order Lagrange tensor), nevertheless, its utility is far from being checked.

Let us observe the characteristic generalized Born-Infeld structure of Lagrangians, which will be our paradigm later on. Another example of generalized Born-Infeld-type models is the simplest one, namely that for the $N$-tuplet of scalar fields on the $n$-dimensional space-time manifold, $n<N$, which analytically represents the scalar field taking values in an $N$-dimensional linear space $W$ endowed with the pseudo-Euclidean or pseudo-Riemannian geometry given by the metric tensor $\eta$ in $U$. If $\phi: M\rightarrow W$ is such a field on the space-time manifold $M$, then it gives rise to the metric-like tensor $\phi^{\ast}\eta$ on $M$, i.e., analytically $g[\phi]_{\mu\nu}=\eta_{AB}(\partial \phi^{A}/\partial x^{\mu})(\partial \phi^{B}/\partial x^{\nu})$. The simplest Born-Infeld-type Lagrangian for $\phi$ is 
$L[\phi]=\sqrt{|{\rm det}\left[g[\phi]_{\mu\nu}\right]|}$.
It may be shown that generalized Born-Infeld-type Lagrangians of this type \cite{Lieb} known also as Chaplygin-type Lagrangians \cite{Ogawa_07} give also reasonable gravitational predictions. This becomes particularly interesting when $W$-space has some special structure which gives rise to some particular metric tensors in $W$. For example, if $W:={\rm L}(U)\simeq U\otimes U^{\ast}$, then
$\eta(X,Y)=\lambda{\rm Tr}(X,Y)+\mu{\rm Tr}X{\rm Tr}Y$
and then
$g[\phi]_{\mu\nu}=\lambda\partial_{\mu}\phi^{A}{}_{B}\partial_{\nu}\phi^{B}{}_{A}+
\mu\partial_{\mu}\phi^{A}{}_{A}\partial_{\nu}\phi^{B}{}_{B}$.
Quite in a similar way, the internal metric $\eta$ may be constructed for more complicated situations when $U$ was a Lie group or the manifold of twice covariant second-order tensors in $U$. In this case the manifold $W$ was endowed in the non-Euclidean Riemannian structure by the metric tensor $\eta$ in $U$. However, there is no place here for the discussion of such problems.

\section{General structure of Lagrangians and their field equations}

Let $\phi=(\ldots,\phi_{A},\ldots): M\rightarrow FM$ be a field of linear frames, i.e., a cross-section of the principal bundle $\pi:FM\rightarrow M$, $\pi\circ\phi={\rm Id}_{M}$. Its dual field of co-frames will be denoted by 
$\widetilde{\phi}=(\ldots,\phi^{A},\ldots): M\rightarrow F^{\ast}M$; 
obviously, we have $\langle\phi^{A},\phi_{B}\rangle=\delta^{A}{}_{B}$. 
The full linear group ${\rm GL}\left(n,\mathbb{R}\right)$ acts on these fields in the sense of the natural action of the structural group on the values of $\phi$ and $\widetilde{\phi}$:
\begin{eqnarray}
\phi&\mapsto&\phi L=(\dots,\phi_{A},\ldots)L=(\dots,\phi_{B}L^{B}{}_{A},\ldots),\label{2.1a}\\
\widetilde{\phi}&\mapsto&\widetilde{\phi}
L=(\dots,\phi^{A},\ldots)L=(\dots,L^{-1A}{}_{B}\phi^{B},\ldots)\label{2.1b}
\end{eqnarray}
for any $L\in{\rm GL}\left(n,\mathbb{R}\right)$. Obviously, $\widetilde{\phi}L$ is identically dual to $\phi L$.

According to the general programme formulated above we search for the following geometric objects built of fields of frames: 1) A first-order Lagrangian, i.e., a differential $n$-form $\mathcal{L}[\phi]$ on $M$. The functional dependence $\phi\mapsto\mathcal{L}[\phi]$ should be first-order local, i.e., $\mathcal{L}$ is to depend pointwise on algebraic values of $\phi$ and of its first derivatives. We require its invariance under ${\rm Diff}(M)$ and under ${\rm GL}^{+}\left(n,\mathbb{R}\right)$:
\begin{equation}\label{2.2}
\mathcal{L}[\varphi^{\ast}\cdot\phi]=\varphi^{\ast}\cdot\mathcal{L}[\phi],\qquad
\mathcal{L}[\phi L]=\mathcal{L}[\phi]
\end{equation}
for any $\phi\in{\rm Diff}(M)$ and $L\in{\rm GL}^{+}\left(n,\mathbb{R}\right)$.
2) A two-fold covariant tensor field $G[\phi]$ on $M$. The dependence
$\phi\mapsto G[\phi]$ should be also first-order local and invariant under ${\rm Diff}(M)$ and ${\rm GL}^{+}\left(n,\mathbb{R}\right)$,
\begin{equation}\label{2.3}
G[\varphi^{\ast}\cdot\phi]=\varphi^{\ast}\cdot G[\phi],\qquad G[\phi L]=G[\phi]
\end{equation}
for any $\varphi\in{\rm Diff}(M)$, $L\in{\rm GL}^{+}\left(n,\mathbb{R}\right)$.

Let us notice that $G[\phi]$ could not be ${\rm GL}^{+}\left(n,\mathbb{R}\right)$-invariant if we restricted ourselves to the algebraic dependence of $G$ on $\phi$, as is commonly done in the metric-teleparallel theories of gravitation \cite{17,22}. Thus, certainly, we cannot put 
$G=\eta_{AB}\phi^{A}\otimes\phi^{B}$, $\|\eta_{AB}\|={\rm diag}(l,-l,\ldots,-l)$.
The object $G[\phi]$ will occur in two roles:
1. $G[\phi]$ is a tool for constructing $\mathcal{L}$, because
$\sqrt{|\det\|G_{ij}\||}$ represents a scalar density of weight one, and we can put
$\mathcal{L}[\phi]=f[\phi]\sqrt{|\det\|G_{AB}\||}\phi^{1}\wedge\ldots\wedge\phi^{n}$,
where $f$ is a ${\rm GL}^{+}\left(n,\mathbb{R}\right)$-invariant scalar built of $\phi$ and of its derivatives, whereas $G_{AB}$ are non-holonomic components of $G$, $G_{AB}=G\left(\phi_{A},\phi_{B}\right)$.
2. $G$ is a candidate for the metric tensor of the physical space-time.
Obviously, if interpreted in this way, $G[\phi]$ must be subject to the symmetry and nonsingularity condition.

As we shall see, the set of natural $G$-s satisfying (\ref{2.3}) is not exhausted by symmetric tensors. A priori it is not clear if $G$ used as a ``brick-stones'' of $\mathcal{L}$ is a proper candidate for the metric tensor of the physical space-time. The invariance requirements, (\ref{2.3}) do not precise $G$ uniquely even within the realm of symmetric tensors. Moreover, once constructed, $\mathcal{L}$ enables us to derive additional expressions compatible with (\ref{2.3}). The proper choice of the metric tensor to be used in Lagrangians of other physical fields can be justified only a posteriori.

To be able to construct $\mathcal{L}[\phi]$ and $G[\phi]$ we have to define an invariant derivative of the field $\phi$. $M$ is a bare manifold without any absolute structure, thus, there is no differentiation of the general tensor fields. Fortunately, the peculiarity of $\phi$ as a field of frames enables us to define its invariant derivative as a system of exterior differentials $d\phi^{A}$, $A=1,\ldots,n$. It is convenient to unify them into a single tensorial object $S$, the torsion tensor of the parallelism. More rigorously: $\phi$ establishes some flat linear connection $\omega[\phi]$ because it is a cross-section of the principal bundle $FM$. The corresponding covariant differentiation is uniquely described by the condition $\nabla\phi_{A}=0$ or, equivalently, $\nabla\phi^{A}=0$. This means that the non-holonomic and holonomic coefficients of the parallelism connection are given by $\Gamma_{\rm tel}[\phi]^{A}{}_{BC}=0$, $\Gamma_{\rm
tel}[\phi]^{i}{}_{jk}=\phi^{i}{}_{A}\phi^{A}{}_{j,k}$.
The torsion tensor has components
$S^{i}{}_{jk}=\Gamma^{i}{}_{[jk]}=
(1/2)\phi^{i}{}_{A}\left(\phi^{A}{}_{j,k}-\phi^{A}{}_{k,j}\right)$,
i.e.,
$S[\phi]=-\phi_{A}\otimes d\phi^{A}=-\phi_{A}\otimes F^{A}$,
$F^{A}{}_{ij}=\phi^{A}{}_{j,i}-\phi^{A}{}_{i,j}$.

The non-holonomic components of $S[\phi]$ with respect to $\phi$ coincide with the anholonomic object of $\phi$ multiplied by the $1/2$-factor:
\begin{equation}\label{2.7}
S[\phi]=\frac{1}{2}\gamma^{A}{}_{BC}\phi_{A}\otimes\phi^{B}\otimes\phi^{C},
\qquad \gamma^{A}{}_{BC}=\langle\phi^{A},\left[\phi_{B},\phi_{C}\right]\rangle,
\end{equation}
i.e.,
$[\phi_{A},\phi_{B}]=\gamma^{C}{}_{AB}\phi_{C}$ and
$d\phi^{A}=(1/2)\gamma^{A}{}_{BC}\phi^{C}\wedge\phi^{B}$.
Obviously, $S$ is ${\rm GL}^{+}\left(n,\mathbb{R}\right)$-invariant:
$S[\phi L]=S[\phi]$, $\gamma^{A}{}_{BC}[\phi
L]=\gamma^{D}{}_{EF}[\phi]L^{-1A}{}_{D}L^{E}{}_{B}L^{F}{}_{C}$.
It is also linear in derivatives and covariant under ${\rm Diff}(M)$,
$S\left[\varphi^{\ast}\cdot\phi\right]=\varphi^{\ast}\cdot S[\phi]$. Thus, it is reasonable to expect that $\mathcal{L}[\phi]$ and $G[\phi]$ will be algebraically constructed of $S[\phi]$. This brings about the question as to the general shape of quantities which can be intrinsically built of $S^{i}{}_{jk}$. It would be rather difficult to answer this question in an exhaustive manner. The following family of algebraic concomitants of $S$ may be easily guessed without any general theory of invariants:
1) covariant $\gamma$-objects of the Killing-Casimir type (all these tensors are symmetric): $\gamma_{i}=2S^{j}{}_{ij}$, $\gamma_{ij}=4S^{k}{}_{im}S^{m}{}_{jk}$,
$\ldots$,
$\gamma_{i_{1}\cdots i_{k}}=2^{k}S^{j}{}_{i_{1}l}S^{l}{}_{i_{2}m}\cdots
S^{p}{}_{i_{k}j}$, $\ldots$;
2) mixed $\Gamma$-objects, thus, $S^{i}{}_{jk}$ itself and
$\Gamma^{i}{}_{jmn}=4S^{i}{}_{jk}S^{k}{}_{mn}$,
$\Gamma^{i}{}_{jkrs}=8S^{i}{}_{mn}S^{m}{}_{jk}S^{n}{}_{rs}$, $\ldots$.
All these objects are skew-symmetric with respect to some pairs of indices. Contracting the indices in the first mixed object we obtain the following skew-symmetric tensor: $\Gamma_{ij}=4S^{k}{}_{lk}S^{l}{}_{ij}=2\gamma_{m}S^{m}{}_{ij}$.

The coefficients $2^k$ are introduced to retain the correspondence with some popular formulas. They are due to the $1/2$-multiplier in (\ref{2.7}). Of course, all these objects are ${\rm GL}\left(n,\mathbb{R}\right)$-invariant and ${\rm Diff}(M)$-covariant. We are especially interested in covariant tensors of the second order and in scalar densities of weight one. The lowest-order tensor objects in the above list are
$\gamma_{i}=2S^{k}{}_{ik}$, $\gamma_{ij}=\gamma_{ji}=4S^{k}{}_{im}S^{m}{}_{jk}$,
$\Gamma_{ij}=-\Gamma_{ji}=4S^{k}{}_{mk}S^{m}{}_{ij}=2\gamma_{m}S^{m}{}_{ij}$,
where $\gamma_{i}$ is linear in derivatives of $\phi$, and $\gamma_{ij}$, $\Gamma_{ij}$ are quadratic functions of derivatives. This leads to the conjecture that the above-defined objects together with $S$ itself (linear in derivatives) are fundamental entities to be used as algebraic brick-stones of $\mathcal{L}$. When constructing Lagrangians, physicists traditionally like the quadratic dependence on derivatives.

The most general second-order real tensor built intrinsically of $S$ is given by
\begin{equation}\label{2.14}
G_{ij}=\lambda\gamma_{ij}+\mu\gamma_{i}\gamma_{j}+\nu\Gamma_{ij},
\end{equation}
where $\lambda$, $\mu$, $\nu$ are real constants. $G$ is homogeneous-quadratic in derivatives. Putting $\nu=0$, we obtain the most general candidate for the metric tensor built intrinsically of $S$:
\begin{equation}\label{2.15}
g_{ij}=\lambda\gamma_{ij}+\mu\gamma_{i}\gamma_{j}= 4\lambda S^{k}{}_{im}S^{m}{}_{jk}+4\mu S^{k}{}_{ik}S^{m}{}_{jm}.
\end{equation}
The second term in (\ref{2.15}) is algebraically singular, therefore, $\lambda$ must not vanish if $g$ is to be a metric tensor. Thus, for simplicity we shall often put $\lambda=1$, i.e., $g_{ij}=\gamma_{ij}+\mu\gamma_{i}\gamma_{j}$.

The first term in (\ref{2.15}), i.e., the main part of $g$, has a characteristic Killing-like structure known from the theory of Lie algebras. This becomes even more evident if we use non-holonomic coefficients. Indeed, (\ref{2.7}) implies that $\gamma_{ij}=\gamma_{AB}\phi^{A}\otimes\phi^{B}$, $\gamma_{AB}=\gamma^{C}{}_{AD}\gamma^{D}{}_{BC}$. If $M$ is a Lie-group underlying manifold and if $\phi=\left(\ldots,\phi_{A},\ldots\right)$ is an ordered basis of the algebra of left-invariant (or right-invariant) vector fields, then $\gamma^{C}{}_{AB}$ are structure constants and $\gamma_{AB}$ are $\phi$-components of the Killing tensor of the algebra. In this case the coefficients $\gamma_{AB}$ are constant. Obviously, in general they are non-constant, i.e., the tensor field $\gamma[\phi]$ is not covariantly constant under the $\phi$-parallelism (the flat connection $\omega[\phi]$ is not $\gamma[\phi]$-metrical).

The way from $\phi$ to $g[\phi]$ is very natural and, as a matter of fact, almost canonical (excepting the arbitrariness of $\lambda$ and $\mu$). It is not so in the case of the relationship between tetrads and metric tensors, used in the tetrad formulation of the conventional gravitation theory and in its metric-teleparallel generalizations, $g=\eta_{AB}\phi^{A}\otimes\phi^{B}$, $\|\eta_{AB}\|={\rm diag}\left(1,-1,\ldots,-1\right)$.
The last expression is less economic because it a priori involves an additional primitive element $\eta$, logically independent of $\phi$. The introducing of $\eta$ is not physically justified if $\phi$ is to be an actually fundamental physical field, the more so, because it destroys the symmetry of degrees of freedom reducing it from ${\rm GL}\left(n,\mathbb{R}\right)$ to ${\rm SO}\left(1,n-1;\mathbb{R}\right)$.

We can formally admit complex constants $\lambda$, $\mu$, $\nu$ in (\ref{2.14}). It seems that one can expect some physical applications in the case of Hermitian tensors, $G_{ij}=G^{\ast}_{ji}$. This corresponds to the choice ${\rm Im}(\lambda)=$ ${\rm Im}(\mu)=0$, ${\rm Re}(\nu)=0$. Thus, the most general Hermitian tensor of the valence $(0,2)$, homogeneous-quadratic in derivatives and built in an intrinsic ${\rm GL}^{+}\left(n,\mathbb{R}\right)$-invariant way from $\phi$, has the form: $K_{ij}=\lambda\gamma_{ij}+\mu\gamma_{i}\gamma_{j}+i\nu\Gamma_{ij}$, where $\lambda$, $\mu$, $\nu$ are real constants.

We shall use the common symbol $T$ for the tensors $G$, $K$. These tensors enable us to construct the natural densities of weight one, namely $\sqrt{|\det\|T_{ij}\||}$. We shall also use the abbreviation $\sqrt{|T|}$. They may be used as Lagrangians. Though every Lagrangian by its very nature is a scalar density of weight one. This fact is often obscured by some extra geometry which enables one to factorize the Lagrange density into the product of some geometrically distinguished (and often forgotten) standard density and the scalar "Lagrangian" which gives an account of the dynamics. The $\sqrt{|T|}$ is the most convenient Lagrangian. The most general ${\rm GL}^{+}\left(n,\mathbb{R}\right)$-invariant Lagrangian densities will be sought as products of the above ``geometrical'' densities and
``dynamical'' scalar multipliers $f_{T}(S)$ built in an algebraic way of the tensor $S$ alone, $L_{T}=f_{T}(S)\sqrt{|T|}$. The particular choice of $T$ within the $3$-parametric families of $G$'s and $K$'s does not matter, since it merely modifies the shape of the function $f_{T}$. Using differential forms we can write
\begin{equation}\label{2.19}
\mathcal{L}=f_{T}(S)\sqrt{|\det\|T_{AB}\||}\phi^{1}\wedge\ldots\wedge\phi^{n},
\end{equation}
where $T_{AB}$ denotes $G_{AB}$ or $K_{AB}$ and
$G_{AB}=\lambda\gamma_{AB}+\mu\gamma_{A}\gamma_{B}+\nu\Gamma_{AB}$,
$K_{AB}=\lambda\gamma_{AB}+\mu\gamma_{A}\gamma_{B}+i\nu\Gamma_{AB}$,
$\gamma_{AB}=\gamma^{C}{}_{AD}\gamma^{D}{}_{BC}$, $\gamma_{A}=\gamma^{B}{}_{AB}$,
$\Gamma_{AB}=\gamma_{C}\gamma^{C}{}_{AB}$.
Expressing (\ref{2.19}) in terms of a local chart $\left(U,x^{1},\ldots,x^{n}\right)$ oriented compatibly with $\phi$, we obtain
$\mathcal{L}|_{U}=f_{T}(S)\sqrt{|T|}dx^{1}\wedge\ldots\wedge dx^{n}$.
Obviously, such Lagrangians satisfy the invariance requirements (\ref{2.2}). The most convenient choice seems to be $T=g$, or simply $T=\gamma$. The action functional is then given by
\begin{equation}\label{2.22}
I\left[\phi,\Omega\right]=\int_{\Omega}f(S)d\mu_{g[\phi]},
\end{equation}
where $\mu_{g[\phi]}$ denotes the pseudo-Riemannian measure induced by $g$.
In general $f$ is a function of some basic scalars built of $S$. Let us mention a few typical examples:
$I_{1}=\gamma_{il}\gamma^{jm}\gamma^{kn}S^{i}{}_{jk}S^{l}{}_{mn}$,
$I_{3}=(1/4)\gamma^{ij}\gamma_{i}\gamma_{j}=
\gamma^{ij}S^{k}{}_{ik}S^{m}{}_{jm}$.
These quantities are affine counterparts of the familiar Weitzenb\"{o}ck invariants $J_{1}$, $J_{3}$ \cite{10,17,22}. The invariant $J_{2}$ has no analogue because we have that $\gamma^{ij}S^{m}{}_{in}S^{n}{}_{jm}=n/4=$ const.
There are also ${\rm GL}\left(n,\mathbb{R}\right)$-invariant scalars of the form
${\rm Tr}\left(\widehat{\Gamma}^{p}\right)=
\Gamma^{i}{}_{j}\Gamma^{j}{}_{k}\cdots\Gamma^{l}{}_{m}\Gamma^{m}{}_{i}$ ($p$ factors).

Let us notice that the above invariants are homogeneous of degree zero in $S$. Thus, $L=f\sqrt{|T|}$ is homogeneous of degree $n$ in $S$ (and consequently, homogeneous of degree $n$ in derivatives of $\phi$). The particular shape of $f$ should be guessed or postulated on the basis of some intuitive or physical reasoning. Then, the choice should be verified by comparing its consequences with experimental data or with some well-established theories, e.g., with the conventional Einstein relativity. From the purely computational point of view,
the simplest model is $f=1$, $T=\gamma$, i.e., the Lagrangian density of the form $L=\sqrt{|\det\|\gamma_{ij}\||}$. Another simple candidates for $L$, next in the order of computational complexity, are given by the square roots of determinants of $G$ and $K$. The characteristic square-root expressions and determinants make these models formally similar to those used in the Born-Infeld electrodynamics \cite{24}. We suppose that there are also physical reasons for this analogy, which perhaps make it something more than a merely formal similarity. Namely, the presented model is some kind of a nonlinear $n$-electrodynamics involving $n$ kinds of ``electromagnetic fields'' with potentials $\phi^{A}$ and field strengths $F^{A}=d\phi^{A}$. There exist $n$ kinds of ``photons'', i.e., quanta of the fields $\phi^{A}$. It is not excluded that the genuine Maxwellian photon has something to do with the field $\phi$.

Another simple Lagrangian of this kind is given by a linear combination of the square-root expressions involving separately the symmetric and antisymmetric tensors: $L=\sqrt{|\det\|4\lambda S^{a}{}_{ib}S^{b}{}_{ja}+4\mu S^{a}{}_{ia}S^{b}{}_{jb}\||}
+\xi\sqrt{|\det\|4\lambda S^{b}{}_{ab}S^{a}{}_{ij}\||}$. However, let us observe that field equations would be irrational then. It is a dark feature of them.

We shall now discuss the general form of the field equations and invariance principles. The action functional will be represented in the form (\ref{2.22}) where we put $g=\gamma$, i.e., we have $\mathcal{L}=f(S)\sqrt{|\det\|\gamma_{AB}\||}\phi^{1}\wedge\ldots\wedge\phi^{n}$,
$L=f(S)\sqrt{|\det\|\gamma_{ij}\||}=f(S)\sqrt{|\det\|4S^{a}{}_{ib}S^{b}{}_{ja}\||}$.
It will be convenient to use the auxiliary tensorial quantity
$\Sigma_{ijk}:=S_{ijk}-S_{jik}=\gamma_{im}S^{m}{}_{jk}-\gamma_{jm}S^{m}{}_{ik}=
-\Sigma_{jik}$.
The quantities $S$ and $\Sigma$ are algebraically equivalent to each other, namely,
$S_{ijk}=\left(\Sigma_{ijk}-\Sigma_{jki}+\Sigma_{kij}\right)/2$.
We shall also use the family of differential two-forms $\Sigma_{A}$, i.e.,
$\Sigma_{Aij}=\Sigma_{ijk}\phi^{k}{}_{A}$.
Obviously, 
$\Sigma_{A}[\phi L]=\Sigma_{B}[\phi]L^{B}{}_{A}$.
Let us introduce the system of ``field momenta'' $H_{A}$, i.e.,
$H_{A}{}^{ij}:=\partial L/\partial\phi^{A}{}_{ij}$.
Its ${\rm GL}\left(n,\mathbb{R}\right)$-transformation rule is identical with that of $\Sigma_{A}$, i.e., $H_{A}[\phi L]=H_{B}[\phi]L^{B}{}_{A}$.
In the case of ${\rm GL}\left(n,\mathbb{R}\right)$-invariant models, the quantities $H_{A}$ become
$H_{A}{}^{ij}=-2f\sqrt{|\gamma|}\Sigma_{A}{}^{ij}+\sqrt{|\gamma|}
P_{a}{}^{ij}\phi^{a}{}_{A}$, $P_{a}{}^{ij}=\partial f/\partial S^{a}{}_{ij}$.

\noindent{\bf Remark:} components of the antisymmetric tensor $S^{a}{}_{ij}$ are not algebraically independent, thus, the derivative is to be understood in the following sense: $\delta f=P_{a}{}^{ij}\delta S^{a}{}_{ij}$, where $P_{a}{}^{ij}=-P_{a}{}^{ji}$ and the higher-order terms are neglected.

The quantities $H_{A}$ are skew-symmetric tensor densities of weight one. The term ``field momentum'' is used by many authors \cite{6-1,6-2,6-3,7}; it is justified by the mechanical analogy, i.e., $P_{i}=\partial L/\partial\dot{q}^{i}$, and by multi-symplectic formulations of the classical field theory \cite{3,13,14}.
The system $(\ldots,H_{A},\ldots)$ is equivalent to the following ${\rm GL}\left(n,\mathbb{R}\right)$-invariant tensor density of weight one:
$H_{k}{}^{ij}=\phi^{A}{}_{k}H_{A}{}^{ij}=\partial L/\partial S^{k}{}_{ij}$.
In ${\rm GL}\left(n,\mathbb{R}\right)$-invariant dynamical models we have
$H_{k}{}^{ij}=\sqrt{|\gamma|}\left(-2f\Sigma^{ij}{}_{k}+P_{k}{}^{ij}\right)$.
For reasons which later become clear, this quantity will be called the local hyperspin or the local affine spin of $\phi$.

\noindent{\bf Remark:} Let us notice that the quantities $\Sigma^{ij}{}_{k}$ and $H_{k}{}^{ij}$ give rise to the additional symmetric second-order tensor fields, i.e., $E^{aj}{}_{k}E^{bk}{}_{j}$ and $|\gamma|^{-1}H_{i}{}^{aj}H_{j}{}^{bi}$. It is not excluded that these tensors should be combined with $\gamma^{ij}$ when we search for a candidate for the proper metric tensor occurring in Lagrangians for other fields and in macroscopic space-time geometry. 

The system of self-interaction currents is given by
$j^{i}{}_{A}=-\partial L/\partial \phi^{A}{}_{i}$.
These quantities are vector densities of weight one in $M$. Their
${\rm GL}\left(n,\mathbb{R}\right)$-transfor\-mation rule is identical with that of the frame $\phi$, i.e.,
$j^{i}{}_{A}[\phi L]=j^{i}{}_{B}[\phi]L^{B}{}_{A}$.
We shall unify this $n$-tuple of currents into a mixed tensor density of weight one, namely
$j^{a}{}_{b}=j^{a}{}_{A}\phi^{A}{}_{b}=-L\delta^{a}{}_{b}+2H_{d}{}^{ca}S^{d}{}_{cb}$.
This object is closely related to the canonical energy-momentum complex
\begin{equation}\label{2.35}
t^{a}{}_{b}=\phi^{A}{}_{c,b}\frac{\partial L}{\partial \phi^{A}{}_{c,a}}-L\delta^{a}{}_{b},
\end{equation}
namely,
$j^{a}{}_{b}=t^{a}{}_{b}+H_{d}{}^{ac}\Gamma_{\rm tel}{}^{d}{}_{bc}$.
The non-tensorial character of $t^{a}{}_{b}$ is compensated by the connection term. Thus, from some point of view, the quantity $j^{a}{}_{b}$ can be interpreted as the energy-momentum tensor density of the field $\phi$. In any case, there exists a striking formal analogy between (\ref{2.35}) and the formula
\begin{equation}\label{2.37}
j^{a}{}_{b}=2\frac{\partial L}{\partial S^{d}{}_{ca}}S^{d}{}_{cb}-L\delta^{a}{}_{b}.
\end{equation}
Roughly speaking, the usual (non-tensorial) derivative $\phi^{A}{}_{i,j}$ in (\ref{2.35}) is replaced in (\ref{2.37}) by the invariant derivative $S^{a}{}_{ij}$.
If we restrict ourselves to generally-covariant and ${\rm GL}^{+}\left(n,\mathbb{R}\right)$-invariant dynamical models, then
\begin{equation}\label{2.38}
J^{i}{}_{A}=S^{i}{}_{kj}H_{A}{}^{kj},\qquad {\rm i.e.},\qquad
j^{a}{}_{b}=S^{a}{}_{cd}H_{b}{}^{cd}.
\end{equation}
This follows from the fact that any covariant and ${\rm GL}^{+}\left(n,\mathbb{R}\right)$-invariant Lagrangian $L$ is homogeneous of degree $n$ in the tensor variable $S$. Indeed, comparing (\ref{2.37}) and (\ref{2.39}) we obtain
$L\delta^{i}{}_{m}=2S^{k}{}_{jm}H_{k}{}^{ji}-S^{i}{}_{kj}H_{m}{}^{kj}$.
Contracting this equation we obtain
$nL=S^{k}{}_{ij}H_{k}{}^{ij}=S^{k}{}_{ij}\left(\partial L/\partial S^{k}{}_{ij}\right)$,
i.e., exactly the homogeneity condition.
The Euler-Lagrange equations read
\begin{equation}\label{2.39}
H_{A}{}^{ij}{}_{,j}=-J^{i}{}_{A}.
\end{equation}
To obtain an explicitly invariant form, we have to express the usual derivative in terms of the covariant one. We can use both the Levi-Civita connection corresponding to $\gamma[\phi]$ and the $\phi$-parallelism connection. We obtain the equations:
\begin{eqnarray}
{\rm Levi-Civita}:&\ &H_{A}{}^{ij}{}_{;j}=-j^{i}{}_{A},\label{2.40}\\
{\rm Parallelism}:&\ &H_{A}{}^{ij}{}_{|j}=
-j^{i}{}_{A}+H_{A}{}^{kj}S^{i}{}_{kj}-2H_{A}{}^{ij}S^{k}{}_{kj}.\label{2.41}
\end{eqnarray}
Substituting here expressions (\ref{2.38}) we obtain
\begin{equation}\label{2.42}
H_{A}{}^{ij}{}_{;j}=S^{i}{}_{jk}H_{A}{}^{kj},\qquad {\rm i.e.},\qquad
H_{k}{}^{ij}{}_{|j}=-2H_{k}{}^{ij}S^{l}{}_{lj},
\end{equation}
or, writing this in a more suggestive and concise way
\begin{equation}\label{2.43}
\left(\nabla_{j}+2S^{l}{}_{lj}\right)H_{k}{}^{ij}=0.
\end{equation}
The obvious identity $H_{A}{}^{ij}{}_{;ji}=0$ leads to the following continuity equations for the self-interaction currents $j_{A}$:
\begin{equation}\label{2.44}
j^{i}{}_{A;i}=j^{i}{}_{A,i}=0.
\end{equation}

There are obvious similarities and analogies between our model and nonlinear electrodynamics. The covector fields $\phi^{K}{}_{i}$ correspond to the covector potential $A_{i}$ and their differentials $F^{K}{}_{ij}=\phi^{K}{}_{j,i}-\phi^{K}{}_{i,j}$ are analogous to the
electromagnetic field strength $F_{ij}=A_{j,i}-A_{i,j}$ (i.e., to the fields $\overline{E}$, $\overline{B}$). Thus, we are dealing with $n$ kinds of ``electromagnetic fields''. The quantities $H_{K}{}^{ij}$ correspond to the field $H^{ij}=\partial L/\partial A_{i,j}$ produced by sources (i.e., to the fields $\overline{D}$, $\overline{H}$). The self-interaction currents $j^{i}{}_{K}=\partial L/\partial \phi^{K}{}_{i}$ correspond to the electric current $j^{i}=\partial L/\partial A_{i}$ and our field equations (\ref{2.40}) are analogous to the Maxwell equations $H^{ij}{}_{;j}=-j^{i}$.

Our Lagrangian is not invariant under Abelian gauge transformation $\phi^{K}\mapsto\phi^{K}+df^{K}$, thus, to some extent it resembles the models used in the electrodynamics of Mie and Proca. Instead of the functional gauge group $\phi^{K}\mapsto \phi^{K}+df^{K}$ ``parameterized'' by $n$ arbitrary functions $f^{K}:M\rightarrow \mathbb{R}$, we have the symmetry group ${\rm Diff}(M)$, also parameterized by a system of $n$ functions $\overline{x}^{i}\left(x^{1},\ldots,x^{n}\right)$, $i=1,\ldots,n$. Nevertheless, our ``photons'', i.e., quanta of $\phi$, are massless because $L$ does not involve any term built algebraically of $\phi$ alone. The currents $j_{K}$ do not correspond to external sources; they are self-interaction currents resulting from the nonlinearity of our models (the field is produced by itself).

The particular shape of our field equations (\ref{2.42}) depends on the factor $f$. In the simplest model, when $f=1$, we have
\begin{equation}\label{2.45}
\Sigma_{A}{}^{ij}{}_{;j}=-\Sigma_{A}{}^{jk}S^{i}{}_{jk},\qquad {\rm i.e.},\qquad
\left(\sqrt{|\gamma|}\Sigma_{A}{}^{ij}\right)_{|j}=
-2\sqrt{|\gamma|}S^{k}{}_{kj}\Sigma_{A}{}^{ij}.
\end{equation}

The field of frames occurs in our model as a fundamental physical quantity, whereas the metric tensor $g[\phi]$ is a secondary object built of $\phi$ and of its first-order derivatives. It is interesting, however, that the same field equations (\ref{2.42}) and the formula (\ref{2.15}) for $g[\phi]$ may be obtained from some variational principle using $\phi$ and $g$ as independent dynamical quantities. The corresponding action functional is given by
\begin{equation}\label{2.46}
I\left[\phi,g|_{\Omega}\right]=
\int_{\Omega}f(S)\left(g^{ij}\gamma_{ij}+2-n\right)\sqrt{|\gamma|}
dx^{1}\wedge\ldots\wedge dx^{n}.
\end{equation}
Variation of this action with respect to $g$ gives $g=\gamma$, and then, varying $\phi$, we obtain (\ref{2.42}). The variational principle based on (\ref{2.46}) and the field equations (\ref{2.45}) (corresponding to the choice $f=1$) will give us some hints concerning the hypothetical relationship between our ${\rm GL}\left(n,\mathbb{R}\right)$-invariant models and the Einstein theory.

The ${\rm GL}^{+}\left(n,\mathbb{R}\right)$-invariance of our model gives rise to $n^{2}$ conserved physical quantities represented by the following differential forms on $M$ (depending on sections $\phi:M \rightarrow FM$): $F^{A}{}_{B}[\phi]=\sum_{j}(-1)^{j-1}F^{A}{}_{B}{}^{j}
dx^{1}\wedge\ldots\wedge_{j}\ldots\wedge dx^{n}$ ($dx^{j}$ dropped out in the $j$-th term), where
$F^{A}{}_{B}{}^{j}=-\phi^{A}{}_{i}H_{B}{}^{ij}=-\phi^{A}{}_{a}\phi^{b}{}_{B}H_{b}{}^{aj}=
-\phi^{A}{}_{a}\phi^{b}{}_{B}\partial L/\partial S^{b}{}_{aj}$. If $\phi$ satisfies equations of motion then
\begin{equation}\label{2.49}
d F^{A}{}_{B}=0, \qquad {\rm i.e.}, \qquad F^{A}{}_{B}{}^{j}{}_{;j}=F^{A}{}_{B}{}^{j}{}_{,j}=0.
\end{equation}
For any pair of numerical indices $A, B$ the quantities $F^{A}{}_{B}{}^{i}$ are components of a vector
density of weight one. The object $F$ obeys the adjoint transformation rule under ${\rm GL}(n, \mathbb{R})$:
$F^{A}{}_{B}[\phi L]=L^{-1 A}{}_{C}F^{C}{}_{D}[\phi]L^{D}{}_{B}$.

To interpret the quantities $F^{A}{}_{B}$ in geometric terms we have to use the natural multi-symplectic
structure induced by $L$ on the bundle $J(\pi)$ of first-order jets of $FM$ over $M$ \cite{3,13,14}. Namely,
$F^{A}{}_{B}$'s give rise to certain differential $(n-1)$-forms $\mathbb{F}^{A}{}_{B}$ on $J(\pi)$,
$F^{A}{}_{B}[\phi]=(j\phi)^{*}\mathbb{F}^{A}{}_{B}$. 
These forms are Hamiltonian generators of the action of 
${\rm GL}(n,\mathbb{R})$ on dynamical variables. Their Poisson brackets are given by
$\{\mathbb{F}^{A}{}_{B}, \mathbb{F}^{C}{}_{D}\}=\mathbb{F}^{C}{}_{B}\delta^{A}{}_{D}-
\mathbb{F}^{A}{}_{D}\delta^{C}{}_{B}$; the coefficients on the right hand side coincide with the structure constants of ${\rm GL}(n,\mathbb{R})$.
Therefore, it is reasonable to call the quantities $F^{A}{}_{B}$ the ``co-moving'' components of the
hyperspin (affine spin) of the field $\phi$. By ``co-moving'' components we mean ``projections'' onto vectors of
the frame $\phi$. The holonomic components $H_{a}{}_{bj}$ fail to be divergence-free, thus, they do not
describe conserved physical quantities. 

To any $(n-1)$-dimensional oriented surface $\Sigma$ we can attribute the integral
$F^{A}{}_{B}[\Sigma]=\int_{\Sigma}F^{A}{}_{B}$.
If $\Sigma$ is a boundary of some regular region and $\phi$ satisfies our field equations, then $F^{A}{}_{B}[\Sigma]=0$, in virtue of (\ref{2.49}). If $\Sigma_{1}$ and $\Sigma_{2}$ are two regular surfaces with the common boundary $\partial \Sigma_{1}=\partial \Sigma_{2}$ and $\phi$ obeys the field equations, then, after using the appropriate convention concerning orientation we have 
\begin{equation}\label{2.50}
F^{A}{}_{B}[\Sigma_{1}]=F^{A}{}_{B}[\Sigma_{2}],
\end{equation}
i.e., the global conservation law corresponding to (\ref{2.49}). If $\gamma$ is
normal-hyperbolic and $\Sigma_{1}, \Sigma_{2}$ are two disjoint space-like surfaces approaching each other at
spatial infinity, then (\ref{2.50}) expresses the time-independence of the total affine spin.
Obviously, instead of assuming $\Sigma_{1}$ and $\Sigma_{2}$ to approach each other at spatial infinity, we can
assume that the field, behave in a proper way, i.e., there is no ``radiation''. For any spatial
section $\Sigma$ we have then $F^{A}{}_{B}[\Sigma]=K^{A}{}_{B}$, where $K^{A{}_{B}}$ are fixed constants characterizing a given solution. The matrix $||K^{A}{}_{B}||$ is some kind of an $L(n, \mathbb{R})$-valued 
``charge'' of $\phi$. It obeys the adjoint transformation rule
$K^{A}{}_{B}[\phi L]=L^{-1 A}{}_{C}K^{C}{}_{D}[\phi]L^{D}{}_{B}$.
This ``charge'' (affine spin) can be characterized in a ${\rm GL}(n,\mathbb{R})$ -invariant way by a system of
eigenvalues of $||K^{A}{}_{B}||$. These are conserved scalars invariant under ${\rm GL}(m,\mathbb{R})$.

Let us now discuss briefly a few consequences of the general covariance, i.e., of the
invariance under the group ${\rm Diff}(M)$. In these considerations we do not assume our models to be
invariant under ${\rm GL}^{+}(n, \mathbb{R})$. The group ${\rm Diff}(M)$ involves arbitrary functions, thus, the required 
general covariance implies a system of differential identities \cite{31,32}. They read 
\begin{eqnarray}
&&\frac{\partial L}{\partial \phi^{A}{}_{b,a}}=-\frac{\partial L}{\partial \phi^{A}{}_{a,b}}, \qquad  {\rm i.e.,} \qquad H_{A}{}^{ba}=-H_{A}{}^{ab},\label{2.51}\\
&&t^{k}{}_{l,k}=\phi^{A}{}_{a,l}\mathcal{L}^{a}{}_{A},\qquad
j^{i}{}_{A}=-\frac{\partial L}{\partial \phi^{A}{}_{i}}=\left(t^{i}{}_{j}-\frac{\partial L}{\partial \phi^{B}{}_{k,i}} \phi^{B}{}_{j,k}\right)\phi^{j}{}_{A},\label{2.52-3}
\end{eqnarray}
where $t^{k}{}_{l}$ denotes the canonical energy-momentum complex and $\mathcal{L}^{a}{}_{A}$ is the $\left({}^{a}_{A}\right)$-th Euler-Lagrange term.
Equation (\ref{2.51}) means that any generally-covariant Lagrangian of the field $\phi$ depends on
derivatives $\phi^{A}{}_{l,a}$ through their skew-symmetric parts, i.e., through the exterior differentials $d\phi^{A}$. The second equation of (\ref{2.52-3}) is equivalent to (\ref{2.37}). Equations (\ref{2.51}), (\ref{2.52-3}) imply that, just as in any generally-covariant theory, the energy-momentum complex is a curl modulo Euler-Lagrange terms, $t^{l}{}_{k}=(D/Dx^{a})H^{la}{}_{k}-\phi^{A}{}_{k}\mathcal{L}^{l}{}_{A}$.
Hence, we have the following strong conservation laws and ``Bianchi identities'':
\begin{equation}\label{2.55}
\frac{D}{Dx^{k}}\left(t^{k}{}_{l}+\phi^{A}{}_{l}\mathcal{L}^{k}{}_{A}\right)=0,
\qquad \frac{D}{Dx^{k}}\left(\phi^{A}{}_{l}\mathcal{L}^{k}{}_{A}\right)
-\phi^{A}{}_{a,l}\mathcal{L}^{a}{}_{A}=0.
\end{equation}
As usual, the weak conservation laws following from the general covariance have the form $(D/Dx^{k})t^{k}{}_{l}=0$.
They are improper, because on realistic motions $t$ becomes a curl \cite{31}. Any one-parameter subgroup of ${\rm Diff}(M)$, i.e., any vector field $u$ on $M$, gives rise to some improper conservation law. The family of all such conservation laws implies continuity equation (\ref{2.44}) for self-interaction currents; nevertheless, (\ref{2.44}) is not a weak conservation law in the literal sense. This is a characteristic feature of any generally-covariant theory whose dynamical variables include
vector fields. Indeed, invariance of $\mathcal{L}$ under the one-parameter group of a vector field $u$: $M \rightarrow TM$ leads to the following identity:
 \begin{equation}\label{2.58}
\frac{D}{Dx^{a}}\left(-t^{a}{}_{b}u^{b}- \frac{L}{D\phi^{B}{}_{l,a}}\phi^{B}{}_{k}u^{k}{}_{,l}\right)=0.
\end{equation}
In principle, this equation belongs to the class of improper conservation laws. However,
after deriving (\ref{2.58}) we can ``forget'' its ``improper'' character and substitute under the
divergence operator the quantities $u^{b}=\phi^{b}{}_{A}$. The resulting laws are exactly the
continuity equations $j^{a}{}_{A,a}=j^{a}{}_{A;a}=0$. Roughly speaking, the conservation law of $j_{A}$ is a
consequence of the invariance of $\mathcal{L}$ under the one-parameter group generated by $\phi_{A}$
($\Gamma[\phi]$-parallel transports in $J(\pi)$ along the direction of $\phi_{A}$).

The vector density $j^{A}$ gives rise, in a standard way, to a differential $(n-1)$-form on $M$,
namely, $\mathcal{J}_{A}[\phi]=\sum_{i}(-1)^{i-1}j^{i}{}_{A}[\phi]dx^{1}\wedge \cdots \wedge_{i} \cdots \wedge dx^{n}$
($dx^{i}$ dropped out in the $i$-th term). Continuity equations (\ref{2.44}) are equivalent to $\mathcal{J}_{A}=0$. The resulting global conservation law tells us that the quantity $J_{A}[\Sigma]:=\int_{\Sigma}{\mathcal{J}_{A}}$
depends on the manifold $\Sigma$ only through its boundary $\partial \Sigma$. This leads in a usual way to the
time-independence of the global charges $Q_{A}$ (if $\phi$ satisfies the field equations). $Q_{A}$ is then
obtained by integrating $\mathcal{J}_{A}$ over any space-like section (Cauchy surface) of $(M, \gamma)$, 
$Q_{A}[\phi]=\int_{\Sigma}\mathcal{J}_{A}[\phi]$. 
Obviously, $Q_{A}[\phi L]=Q_{B}[\phi]L^{B}{}_{A}$. 

In contrast to the currents  $j_{A}$, the mixed quantity $j^{a}{}_{b}$ (the
``energy-momen\-tum'' of $\phi$) is not conserved. Interpretation of $j^{a}{}_{b}$ as a corrected (tensorial)
energy-momentum density of $\phi$ suggests us to interpret the charges $Q_{A}$ as co-moving components
of the total energy-momentum. This interpretation is supported by the
fact that the conservation of $Q_{A}$ is equivalent to the dynamical invariance of our model under
translations along $\phi_{A}$. If $\phi_{A}$ is time-like in $(M,y)$, and $\phi_{B}{}$'s with $B\neq A$ are space-like, then it is quite natural to interpret the formula for $Q_{A}$ as a summation of rest energies of
infinitesimal portions of the physical system described by $\phi$. Conservation of $Q_{A}$ has to do with
functionally-parameterized groups of symmetries, thus for smooth, nonsingular solutions well-behaving at
spatial infinity, $Q_{A}{}$'s will vanish.

Let us now consider a physical system which, in addition to the field of frames, involves
other fields. How to modify our Lagrangians and field equations? We shall not try to discuss
this problem in an exhaustive manner and in all its mathematical generality. At this stage it
is sufficient to restrict ourselves to a few natural hypotheses and qualitative comments.

If $\psi$ is any bosonic field, then the system $(\phi, \psi)$ can autonomously exist in a bare manifold, i.e., it admits a first-order ${\rm Diff}(M)$-invariant Lagrangian. It is apparently natural to put it, following Hilbert/Einstein, in the following form:
\begin{equation}\label{2.60}
\mathcal{L}[\phi, \psi]=\mathcal{L}_{\rm pr}[\phi]+\mathcal{L}_{\rm ass}[\phi, \psi],
\end{equation}
where $\mathcal{L}_{\rm pr}[\phi]$ is a Lagrangian of the pure field $\phi$; the symbol ``pr'' refers to the principal
bundle. The second term of $\mathcal{L}$, involving $\psi$, is denoted by $\mathcal{L}_{\rm ass}$ because $\psi$ is a cross-section of an associated bundle of $FM$. The field $\psi$ can be interpreted as a ``matter'' injected into $M$. 

The simplest reasonable model of $\mathcal{L}_{\rm ass}$ is that quadratic in $\psi$. To construct it we have to use some metric field on $M$. If $\mathcal{L}[\phi, \psi]$ is to be ${\rm GL}^{+}(n, \mathbb{R})$-invariant, then we should use $\gamma[\phi]$ (or, more generally, $g[\phi]$). For example, if $\psi$ is a scalar field, then the simplest choice is
$\mathcal{L}_{\rm ass}[\phi, \psi]=f_{\rm ass}[\phi, \psi]\sqrt{|{\det} ||\gamma_{AB}||\ |}\phi^{1} \cdots \phi^{n}$, i.e.,
 \begin{equation}\label{2.61}
L_{\rm ass}[\phi, \psi]=f_{\rm ass}[\phi, \psi]\sqrt{|{\det} ||\gamma_{ij}||\ |}, \qquad f_{\rm ass}=\frac{1}{2}\gamma^{ij}\psi_{,i}\psi_{,j}-\frac{m^{2}}{2}\psi^{2}.
\end{equation}
Field equations will be linear in $\psi$ but strongly nonlinear in $\phi$. The coupling between
``geometrical'' and ``physical'' quantities (between $\phi$ and $\psi$) has a much more complicated structure than the corresponding coupling between $g$ and $\psi$ in Einstein theory. Indeed, quite independently of the nature of $\psi$, even for scalar fields and covector fields (Proca fields), the variational derivative $\delta L_{\rm ass}/\delta \phi$
involves the second derivative of both $\phi$ and $\psi$. The same is true for $\delta L_{\rm ass}/\delta \psi$. In other words, the subsystems
$\delta L/\delta \psi=0$, $\delta L/\delta \phi=0$
are coupled through second derivatives. Thus, the mutual interaction between ``geometrical''
and ``physical'' degrees of freedom is very essential. Thus, from dynamical viewpoint, it is
rather artificial to distinguish ``geometrical'' and ``physical'' subsystems of the total system of field equations. 

Let us notice that the interaction model (\ref{2.60}) is not very natural in spite of its formal similarity to the Einstein-Hilbert procedure. It is taken almost
literally from the Einstein theory and from the metric-teleparallel theories of gravitation.
Its advantage is that it provides a universal pattern for coupling all tensor fields on $M$
through the field $\phi$. However, this advantage is rather illusory in view of the aforementioned
coupling of subsystems through second derivatives. This coupling occurs even in simplest
and most natural models with $\mathcal{L}_{\rm ass}$ depending on $\phi$ only through the metric 
tensor $\gamma[\phi]$ and the parallel connection $\Gamma[\phi]$. Thus, there is no motivation
at all for preferring the artificial splitting (\ref{2.60}). In ${\rm GL}^{+}(n, \mathbb{R})$-invariant 
theories of the field of frames this artificial structure of $\mathcal{L}$ would be the price paid
for nothing. On the contrary, let us recall that, in metric-teleparallel theories of gravitation 
(including Einstein theory) it is just the peculiar and artificial model
$L=R\sqrt{|g|}+L_{\rm mat}[g, \psi]$ 
(with
$L_{\rm mat}$ depending on $g$ algebraically and through the Levi-Civita covariant derivatives of $\psi$)
which enables us to avoid the coupling through second derivatives (for realistic fields $\psi$).
Mathematically artificial becomes physically privileged. This does not seem to be the case in
${\rm GL}^{+}(n, \mathbb{R})$-invariant models.

In our opinion, in ${\rm GL}^{+}(n, \mathbb{R})$-invariant theory of the field of frames $\phi$ interacting with the
scalar field $\psi$, it would be rather natural to replace (\ref{2.61}) by the Born-Infeld-type Lagrangian density, i.e.,
\begin{equation}\label{2.62}
L=\sqrt{|{\det} ||(1+\mu \psi^{2})\gamma_{ij}+\lambda \psi_{,i}\psi_{,j}|| \ |},
\end{equation}
where $\lambda$, $\mu$ are real constants. The parameter $\mu$ would be responsible
for the mass of the scalar field $\psi$. If $\lambda$ and $\mu$ approach zero then (\ref{2.62}) asymptotically
becomes (\ref{2.61}). Obviously, we can complicate (\ref{2.62}) by allowing $\lambda$ and $\mu$ to depend on $\psi$, or
by multiplying the total expression by a dynamical factor $f(S, \psi, \nabla \psi)$. The system of
scalar invariants from which $f$ can be built is now much richer than in the case of the field
$\phi$ alone. Indeed, besides of $I_{1}$ and $I_{3}$ we can take, e.g., $\psi^{2}$, $\gamma^{ij}\psi_{,i}\psi_{,j}$, etc. In spite of the irrational structure of $L$, the resulting field equations are rational; the same holds for (\ref{2.64}) and for any Born-Infeld-type Lagrangian.
If we consider $\phi$ interacting with the vector field $A$, then we can construct $L$ from the
scalar quantities of the form
$\alpha_{k}=A^{i_{1}}{}_{i_{2}}A^{i_{2}}{}_{i_{3}} \cdots A^{i_{k-1}}{}_{i_{k}} A^{i_{k}}{}_{i_{1}}$, $\beta=\gamma_{ij}A^{i}A^{j}$, etc.
For instance, we can put
\begin{equation}\label{2.64}
L=f(I_{1},I_{3},\alpha \cdots \alpha_{r},\beta)\sqrt{|{\det}||\gamma_{ij}+\mu A_{i}A_{j}+\lambda A_{[i,j]}||\ |},
\end{equation}
where $A$ and $\mu$ are real constants; $\mu$ is responsible for the mass of the field $A$. Obviously, the simplest model is that corresponding to $f=1$.

Some interesting ideas concerning the tetrad-matter interaction are mentioned in the papers of P. Godlewski \cite{PG_02,PG_03_1,PG_05,PG_08,PG_10,PG_13}.
For comparison reasons let us only mention about the generally covariant tetrad models invariant also under the internal Lorentz group ${\rm SO}(1,n-1)\subset{\rm GL}^{+}(n,\mathbb{R})$. In commonly used models one takes as basic quantities the following three Wietzenb\"ock invariants:
$\mathcal{J}_{1}=h_{\alpha\mu}h^{\beta\nu}h^{\gamma\varkappa}S^{\alpha}{}
_{\beta\gamma}S^{\mu}{}_{\nu\varkappa}$,
$\mathcal{J}_{2}=h^{\mu\nu}S^{\alpha}{}_{\mu\beta}S^{\beta}{}_{\nu\alpha}$,
$\mathcal{J}_{3}=h^{\mu\nu}S^{\alpha}{}_{\alpha\mu}S^{\beta}{}_{\beta\nu}$.
One can show that the curvature scalar density $R[h]\sqrt{|h|}$ may be expressed as follows:
\begin{equation}\label{3.10}
R[h]\sqrt{|h|}=\left(\mathcal{J}_{1} +2 \mathcal{J}_{2}-4\mathcal{J}_{3}\right)\sqrt{|h|}+
4 \stackrel{(h)}{\nabla}_{\mu} \left(S^{\alpha}{}_{\alpha\beta}h^{\beta\mu}\sqrt{|h|}\right).
\end{equation}
Therefore, the Hilbert-Einstein Lagrangian (\ref{3.10}) differs from the standard expression
$L_{\rm HE}=\left(\mathcal{J}_{1} +2 \mathcal{J}_{2}-4\mathcal{J}_{3}\right)\sqrt{h}$
by the divergence term given by the following expression: $4\stackrel{(h)}{\nabla}_{\mu}(S^{\alpha}{}_{\alpha\beta}h^{\beta\mu}\sqrt{|h|})=4 (S^{\alpha}{}_{\alpha\beta}h^{\beta\mu}\sqrt{|h|})_{,\mu}$. 
There were plenty of attempts to manipulate with the structure of theory in terms of coefficients at $\mathcal{J}_{1}$, $\mathcal{J}_{2}$, $\mathcal{J}_{3}$, or just by replacing $L_{\rm HE}$ by some general function of invariants. But then the local ${\rm SO}(1,n-1)$-invariance is lost. But when accepting this, there are no more obstacles against replacing ${\rm SO}(1,n-1)$ by the globally acting ${\rm GL}(n,\mathbb{R})$.

\section{Special solutions: semisimple Lie groups with trivial central extensions}

At this rather general stage we have no convincing criteria for any choice of the dynamical factor $f$ occurring in the affinely-invariant models (\ref{2.19}). Thus, it is rather hard to say anything about rigorous solutions of our field equations. Moreover, a priori we do not know at all whether these equations are consistent or not (it is not difficult to formulate artificial variational principles leading to contradictory field equations). We shall not discuss the integrability problem in a systematic way; instead we shall try to construct explicitly some intuitive special solutions independent of the choice of $f$. In conventional Einstein relativity there exist such a priori evident solutions, namely, those corresponding to the flat space-times. The same is true in metric-teleparallel theories of gravitation based on quadratic Lagrangians 
\begin{equation}\label{3.11}
L=(c_{1}\mathcal{J}_{1} +c_{2}\mathcal{J}_{2}+c_{3}\mathcal{J}_{3})\sqrt{|h|}=
c_{1}L_{1}+c_{2}L_{2}+c_{3}L_{3},
\end{equation} 
where quite independently of the assumed dynamical model (coefficients $c_{i}$ in (\ref{3.11})) any holonomic field of frames is a solution. Indeed, if $\phi$ is holonomic, then $S[\phi]=0$ and the field equations resulting from $\rm L$ (\ref{3.11}) are satisfied (they are linear in $S$), the corresponding pseudo-Riemannian manifold $(M,h[\phi]$]) is flat. The fields $\phi_{A}$, $A=1,\ldots,n$ span an Abelian Lie algebra of vector fields, i.e., they generate a local commutative Lie group of transformations; the local action of this group on $M$ is free and transitive. The existence of such solutions is a characteristic feature of the quadratic metric-teleparallel models; it seems to be a natural consequence of the
restriction of ${\rm GL}^{+}(n,\mathbb{R})$ symmetry to ${\rm SO}(1,n-1;\mathbb{R})$. 

\noindent{\bf Remark:} Let us notice that holonomic fields $\phi$ admit adapted charts in which $\phi^{A}{}_{i}$'s are constant. If we consider slightly perturbed fields $\phi'^{A}{}_{i}= \phi^{A}{}_{i} + u^{A}{}_{i}$ then, in the first order of approximation, infinitesimal diffeomorphisms of $M$ result in the following transformation rule for perturbations $u^{A}$:
$u^{A}{}_{i} \mapsto u^{A}{}_{i}+ f^{A}{}_{,i}$, where $f^{A}{}_{,s}$ are scalars. This resembles the gradient gauge rule for covector fields. The general covariance of rigorous equations leads to the gradient invariance of their Jacobi equations. This possibility of deriving the Abelian gauge invariance from the general covariance (i.e., from the invariance under ${\rm Diff}M$) is interesting in itself and can lead to some reflections and hypotheses.
Unfortunately, such solutions do not exist in ${\rm GL}^{+}(n,\mathbb{R})$-invariant models, because it is obvious from the very beginning that $\phi$ must be non-holonomic; otherwise $g[\phi]$ certainly could not be nonsingular. 

Let us define a field of frames $\phi: M \rightarrow FM$ to be {\it Killing-nonsingular} ({\it K-nonsingular}) if its Killing tensor $\gamma[\phi]$ is non-degenerate.
Obviously, our search for solutions of (\ref{2.43}) must be restricted to the variety of K-nonsingular fields. Holonomic fields of frames satisfy $S=0$, i.e., $\gamma^{A}{}_{BC}=0$. The simplest natural generalization of such fields consists in putting $\gamma^{A}{}_{BC}={\rm const}$. The torsion tensor $S$ is then covariantly constant under the $\phi$-parallelism, $S^{i}{}_{jk \left|z\right.}=0$. From the other side, a field of frames $\phi$ is said to be {\it closed} if it has the above property, i.e., if its torsion tensor $S[\phi]$ is constant under $\phi$-parallel transports, $\nabla S[\phi]=0$. Therefore, if $\phi$ is closed, then its component-fields $\phi_{A}$ span a Lie algebra in the Lie-bracket sense. The corresponding local Lie group of transformations acts freely and transitively in open domains of $M$. The tensor $\gamma[\phi]$ is then covariantly constant with respect to the $\phi$-parallelism connection,
$\nabla \gamma[\phi]=0$, i.e., $\gamma_{ij \left|k\right.}=0$.
The non-holonomic components of $\gamma$, $\gamma_{AB}=\gamma \left(\phi_{A},\phi_{B}\right)= \gamma_{ij}\phi^{i}{}_{A}\phi^{j}{}_{B}$, are constant; they coincide with coefficients of the natural Killing form of the Lie algebra $\oplus_{i=1}^n \mathbb{R} \ \phi_{A}$, $\gamma_{AB}=\gamma^{C}{}_{AD}\gamma^{D}{}_{BC}$.

If $\phi_{A}$ is K-nonsingular and closed, then  
$\mathfrak{g}=\oplus_{i=1}^n \mathbb{R} \ \phi_{A}$ 
is a semisimple Lie algebra. Thus, we obtain a local semisimple Lie group of transformations acting freely and transitively in $M$. If we fix some ``origin'' $e \in M$, then $M$ becomes a local semisimple Lie group, $e$ being its identity element. Linear combinations of the vector fields $\phi_A$ (with constant coefficients) become generators of the left regular translations; obviously, they are right-invariant vector fields on the resulting Lie group. The left-invariant vector fields corresponding to $\phi_{A}$'s will be denoted by $\phi^{*}_{A}$; their linear shell (over reals) generates the group of right regular translations. We have the following system of basic commutators (Lie brackets): $\left[\phi_{A},\phi_{B}\right]= \gamma^{C}{}_{AB} \ \phi_{C}$,
$\left[\phi^{*}_{A},\phi^{*}_{B}\right]= - \gamma^{C}{}_{AB} \ \phi^{*}_{C}$, $\left[\phi_{A},\phi^{*}_{B}\right]= 0$.
Obviously, the fields $\phi^{\ast}_{A}$ depend not only on the original fields $\phi_{B}$, $B=1,\ldots,n$, but also on the choice of the neutral point (``origin'') $e \in M$.

The metric tensor $\gamma[\phi]$ admits at least $2n$-dimensional group of motions, because $\phi_{A}$'s and $\phi^{*}_{A}$'s are Killing vectors (infinitesimal isometries), i.e., $\mathcal{L}_{\phi_{A}} \gamma[\phi]= \mathcal{L}_{\phi^{*}_{A}} \gamma[\phi]=0$. Closed fields of frames provide the simplest Lie-algebraic generalization of holonomic ones and at the same time they do not exclude the required nonsingularity of the Killing tensor $\gamma[\phi]$. Thus, they seem to be a candidate for geometrically privileged solutions of ${\rm GL}^{+}(n,\mathbb{R})$-invariant dynamical models. We have in fact the following theorem: 

\noindent{\bf Theorem:} Any closed Killing-nonsingular field of linear frames is a solution of ${\rm GL}^{+}(n,\mathbb{R})$-invariant field equations (\ref{4.3}) independently of the choice of $f$:
\begin{equation}
\nabla_{j} H_{k}{}^{ij}=-2H_{k}{}^{ij}S^{z}{}_{zj}. \label{4.3}
\end{equation}
{\bf Proof:} In any ${\rm GL}^{+}(n,\mathbb{R})$-invariant model the quantity $H$ is an algebraic function of the torsion tensor $S$. Thus, the parallel invariance of $S$ implies that, $\nabla_{z} H_{k}{}^{ij}=0$, in particular $\nabla_{j} H_{k}{}^{ij}=0$. At the same time $S^{m}{}_{mj}=0$, because for any semisimple Lie algebra the structural constants are traceless, $\gamma^{A}{}_{AB}=0$. Therefore, the both sides of (\ref{4.3}) do vanish, i.e., closed nonsingular frames satisfy our field equations. \rule{5pt}{5pt}

Therefore, invariance under ${\rm GL}^{+}(n,\mathbb{R})$ seems to be responsible for the existence of solutions equivalent to local semisimple Lie groups of transformations acting freely and transitively on $M$. Abelian and semisimple Lie groups are opposite special cases within the family of all Lie groups. In this sense affinely-invariant models and metric-teleparallel models are complementary. Therefore, closed-parallelism solutions of ${\rm GL}^{+}(n,\mathbb{R})$-inva\-riant models seem to be conceptual counterparts of holonomic flat-space solutions in metric-teleparallel theories. Unfortunately, in a four-dimensional space-time there are no solutions of this type, because there are no four-dimensional semi-simple Lie algebras. Thus, if we insist on Lie-algebraic solutions as something fundamental, then we must accept Kaluza's philosophy of multidimensional space-times. The ``usual'' four-dimensional space-time would be merely some aspect of ``Kaluza's world'' (e.g., a quotient manifold or a submanifold). We could also try to consider some kind of a complexified four-dimensional space time, because there exist eight-dimensional semisimple Lie algebras (e.g., ${\rm SU}(3)$,
${\rm SL}(3,\mathbb{R})$).

However, it is also possible to retain intuitive special solutions of group-theoretical origin without introducing the above-mentioned complications (increase of dimension, Kaluza's universe, etc.). It turns out that dimensions ``semi\-simple plus one'' are also acceptable. Obviously, this covers the physical dimension four, because there are two simple three-dimensional Lie algebras, $so(3,\mathbb{R})= su(2)$, $so(1,2;\mathbb{R}) = sl(2,\mathbb{R})$. We shall now describe those group-theoretical solutions adapted to dimensions ``semisimple plus one''.
The following notational convention will be used: coordinate and tonsorial indices in an $n$-dimensional manifold run from $0$ to $(n-1)$ and are denoted by Latin letters; Greek indices (``spatial'') run from $1$ to $(n-1)$. Non-holonomic indices are denoted, as usual, by capital symbols, with the same convention concerning Latin and Greek types.

Let us begin with an auxiliary field of frames 
$\left(\psi_{0}, \ldots,  \psi_{\Lambda}, \ldots\right)$, where $\Lambda=1, \ldots, (n-1)$, with the following properties:
1) $\psi_{\Lambda}$'s are invariant under $\phi_{0}$, i.e.,
\begin{equation}
\left[\psi_{0},\psi_{\Lambda}\right]=0, \label{4.4}
\end{equation}
2) $\psi_{\Lambda}$'s span an $(n-1)$-dimensional semisimple Lie algebra, i.e.,
\begin{equation}
\left[\psi_{\Lambda},\psi_{\Sigma}\right]=\psi_{\Delta}C^{\Delta}{}_{\Lambda \Sigma}, \label{4.5}
\end{equation}
where $C^{\Delta}{}_{\Lambda \Sigma}$ are constant and the Killing matrix $C_{\Lambda \Sigma}=C^{\Delta}{}_{\Lambda \Pi}C^{\Pi}{}_{\Sigma \Delta}$ is 
nonsingular.
The ``tetrad'' $\left(\ldots,\psi_{A},\ldots\right)=\left(\psi_{0}, \ldots, \psi_{\Lambda}, \ldots \right)$ is a basis of an $n$-dimensional Lie algebra. Obviously, this algebra is not semisimple; it is a direct product of the one-dimensional centre by $\psi_{0}$ and of the $(n-1)$-dimensional semisimple algebra spanned by $\left( \ldots,\psi_{\Lambda},\ldots \right)$. Thus, it is certainly inapplicable as a candidate for a solution of affinely-invariant equations (\ref{4.3}); the corresponding Killing tensor would be singular. However, we can easily construct from $\psi$ some modified fields of frames which are free of these disadvantages and turn out to be solutions of (\ref{4.3}).
Let $\phi = \left(\phi_{0}, \ldots, \phi_{\Lambda}, \ldots \right)$ be a cross-section of $FM$:
\begin{equation}\label{4.6}
(i)\quad \phi_{0}:= \psi_{0},\qquad
(ii)\quad \phi_{\Lambda}:= \psi_{\Sigma} \ \lambda^{\Sigma}{}_{\Lambda},
\end{equation}
where $\lambda: M \rightarrow {\rm GL}(n-1,\mathbb{R})$ is a matrix-valued function on $M$ constant on all $(n-1)$-dimensional integral surfaces of the distribution spanned by $\left(\ldots,\phi_{\Lambda},\ldots \right)$, or, equivalently, by $\left(\ldots,\psi_{\Lambda},\ldots \right)$. Obviously, (\ref{4.5}) implies that this distribution actually is integrable. If $\lambda$ is not constant all over $M$, then neither the $R$-linear span of $\left(\ldots,\phi_{A}, \ldots \right)$ nor that of $\left(\ldots, \phi_{\Lambda}, \ldots \right)$ are Lie algebras; instead of this we have
$\left[\phi_{A}, \phi_{B}\right]=\gamma^{C}{}_{AB} \ \phi_{C}$, where the coefficients $\gamma^{C}{}_{AB}$ are non-constant functions on $M$. Nevertheless, they are constant along all integral surfaces of the distribution spanned by the system of spatial vectors $\left(\ldots \phi_{\Lambda} \ldots \right)$. They can easily be expressed through the structural constants $C$ and deformation matrices $\lambda$, namely,
$\gamma^{O}{}_{\Lambda O}=0$, $\gamma^{\Sigma}{}_{O \Lambda}= (\lambda^{-1} \dot{\lambda})^{\Sigma}{}_{\Lambda}$,
$\gamma^{\Sigma}{}_{\Lambda \Pi}= C^{\Omega}{}_{\Delta \Gamma}\lambda^{\Delta}{}_{\Lambda}\lambda^{\Gamma}{}_{\Pi}\lambda^{-1 
\Sigma}{}_{\Omega}$, where $\dot{\lambda}:= \phi_{O} \cdot \lambda = \psi_{O} \cdot \lambda$. 
The last quantity also is constant on all integral manifolds of the $\left(\ldots, \phi_{\Lambda}, \ldots \right)$-distribution. Coefficients of the Killing object are given by
$\gamma_{OO}=\gamma^{\Sigma}{}_{O \Lambda} \gamma^{\Lambda}{}_{O \Sigma}= {\rm Tr}((\lambda^{-1} \dot{\lambda})^{2})$,
$\gamma_{ \Lambda\Sigma}= \gamma^{\Delta}{}_{\Lambda\Pi}\gamma^{\Pi}{}_{\Sigma\Delta}= C_{\Pi\Delta}\lambda^{\Pi}{}_{\Lambda}\lambda^{\Delta}{}_{\Sigma}$,
$\gamma_{O \Lambda}= \gamma^{\Delta}{}_{O \Pi}\gamma^{\Pi}{}_{\Lambda \Delta}$, where 
$\gamma_{OO}$ is constant along integral surfaces of the ``spatial'' subframe $\left(\ldots \phi_{\Lambda} \ldots \right)$. ``Spatial'' coefficients $\gamma_{\Lambda \Sigma}$ depend only on the ``spatial'' components $\gamma^{\Omega}{}_{\Pi\Delta}$ of the total non-holonomy object $\gamma^{A}{}_{BC}$. They are built of them in the sense of the $(n-1)$-dimensional Killing formula. Moreover, they are $\lambda$-transforms of the Killing form for the $(n-1)$-dimensional Lie algebra spanned by the original fields $\psi_{\Lambda}$. Therefore,
$\gamma_{\Lambda \Sigma} \ \phi^{\Lambda} \otimes \phi^{\Sigma} = C_{\Lambda \Sigma} \ \psi^{\Lambda} \otimes \psi^{\Sigma}$.
In other words: the ``spatial triad'' $\left(\ldots \phi_{\Lambda} \ldots \right)$ ``breathes'' in the course of ``time'' (group parameter) of $\phi_{0}$, nevertheless the corresponding ``spatial metric'' does not feel this breathing and equals the Lie-algebraic Killing expression built of the field $\left(\ldots \psi_{\Lambda} \ldots \right)$.

It is natural to demand $\gamma_{O\Lambda}=0$, i.e., the mutual orthogonality of integral curves of $\phi_{0}$ and integral surfaces of $\phi_{1}\wedge \ldots \wedge \phi_{n-1}$. Let us now recall that the algebra spanned by $\left(\ldots \psi_{\Lambda} \ldots \right)$ is semisimple and the structural constants $C^{\Lambda}{}_{\Sigma \Delta}$ are skew-symmetric in indices $\Lambda$, $\Delta$ (and consequently in indices $\Lambda$, $\Sigma$) with
respect to the Killing tensor
$C^{\Lambda \Pi} \ C_{\Delta \Omega} \ C^{\Omega}{}_{\Sigma \Pi} = -C^{\Lambda}{}_{\Sigma \Delta}$.
This means that the ``spatial'' quantity $\left\| \gamma^{\Pi}{}_{\Lambda \Delta} \right\|$ is skew-symmetric with respect to $\left\| \gamma_{\Omega \Delta} \right\|$, i.e., $\gamma^{\Lambda \Pi} \ \gamma_{\Delta \Omega} \ \gamma^{\Omega}{}_{\Sigma \Pi} = -\gamma^{\Lambda}{}_{\Sigma \Delta}$. Thus, the quantity $\left\| \gamma_{O \Lambda} \right\|$ will vanish if $\left\| \gamma^{\Delta}{}_{O \Pi} \right\|$ will be $\left\| \gamma_{\Omega \Delta} \right\|$-symmetric, $\gamma_{\Lambda \Delta} \ \gamma^{\Sigma \Pi} \ \gamma^{\Delta}{}_{O \Pi} = \gamma^{\Sigma}{}_{O \Lambda}$. It is easy to show that this implies $\dot{\lambda}\lambda^{-1}$ to be $C$-symmetric, $C_{\Lambda \Delta} \ C^{\Sigma \Pi} (\dot{\lambda}\lambda^{-1})^{\Delta}{}_{\Pi } = (\dot{\lambda}\lambda^{-1})^{\Sigma}{}_{\Lambda }$, i.e., $\lambda$ should be purely deformative. The simplest model is a uniform dilatation,
\begin{equation}
\lambda^{\Sigma}{}_{\Delta} = \lambda \ \delta^{\Sigma}{}_{\Delta}, \label{4.15}
\end{equation}
$\lambda: M \rightarrow \mathbb{R}$ being constant on integral surfaces of $\phi_{1}\wedge \ldots \wedge \phi_{n-1}$. From now on we restrict ourselves
to such fields. This choice implies that $\gamma_{OO}=(n-1)(\dot{\lambda}/\lambda)^{2}=(n-1)(\dot{\ln \lambda})^{2}>0$,
and, finally, the Killing tensor $\gamma[\phi]$ has the following form:
\begin{equation}
\gamma = (n-1)\left(\frac{\dot{\lambda}}{\lambda}\right)^{2} \phi^{O} \otimes \phi^{O} + \gamma_{\Lambda \Sigma} \ \phi^{\Lambda} \otimes \phi^{\Sigma},
\label{4.17}
\end{equation}
where the last term
$\underset{(n-1)}{\gamma}[\psi]=\gamma_{\Lambda \Sigma} \ \phi^{\Lambda} \otimes \phi^{\Sigma}= C_{\Lambda \Sigma} \ \psi^{\Lambda} \otimes \psi^{\Sigma}$
describes $(n-1)$-dimensional metric geometry on integral surfaces of $\phi_{1}\wedge \ldots \wedge \phi_{n-1}$. Let us notice that if vectors $\psi_{1}, \ldots, \psi_{n-1}$ span a compact Lie algebra, then $\underset{(n-1)}{\gamma}[\psi]$ is negatively definite and the tensor $\gamma[\phi]$ is automatically normal-hyperbolic, i.e., its signature equals $(+,-, \ldots, -)$.

In the physical case $(n = 4)$ we have at disposal two simple Lie algebras, $so(3, \mathbb{R})= su(2)$ and $so(1,2;\mathbb{R}) = sl(2,\mathbb{R})$. The algebra $so(3, \mathbb{R})$ is compact, thus, the corresponding $\gamma[\phi]$ is normal-hyperbolic, $\phi_{0}$ is time-like and $\phi_{\Sigma}, \psi_{\Sigma}$ are space-like. Maximal integral surfaces of the $(\ldots, \phi_{\Sigma}, \ldots)$-distribution then become spatial sections, $\phi_{0}$ is a reference frame (ether) and in this way the above metaphoric terms ``time'' and ``space'' acquire a literal relativistic meaning. The Killing signature of $so(1,2;\mathbb{R})$ is $(++-)$, thus the total $4$-dimensional $\gamma[\phi]$ again would be normal-hyperbolic with signature $(+++-)$ (the vector $\phi_{0}$ would be space-like this time). However, from the global point of view such a model is useless because the time-like dimension corresponds to the compact subgroup of planar rotations in ${\rm SO}(1,2;\mathbb{R})$. Pseudo-Riemannian manifolds with closed time-like curves are (as yet) unacceptable as realistic models of the physical space-time. 

Fields of linear frames of the form described above (\ref{4.4})--(\ref{4.6}), (\ref{4.15}) are called {\it breathing-closed fields}. A breathing-closed field $\phi$ is said to be proper if the Lie algebra $\bigoplus_{i=1}^{n-1}R\phi_{i}$ is compact, i.e., if $\gamma[\phi]$ is normal-hyperbolic.
Obviously, the fields of frames induced by a proper breathing-closed field $\phi$ on the $(n-1)$-dimensional ``spatial sections'' (maximal integral surfaces of $\phi_{1}\wedge \ldots \wedge \phi_{n-1}$ are closed and the resulting $(n-1)$-dimensional Riemannian structures are locally isometric with the Killing-Riemann geometry of the compact, semisimple and simply connected Lie group determined uniquely by the structural constants $C^{\Lambda}{}_{\Sigma \Omega}$. The restrictions of $(n-1)$-tuples $(\phi_{1}, \ldots, \phi_{n-1})$ to ``spatial sections'' satisfy $(n-1)$-dimensional equations of the form (\ref{4.3}). The last statement is trivial. However, it turns out that the complete $n$-tuple $\left(\phi_{0}, \ldots, \phi_{n-1}\right)$ is a solution of the original $n$-dimensional system. 

\noindent{\bf Theorem:}  Any breathing-closed field of linear frames satisfies affinely-invariant equations (\ref{4.3}) with all possible dynamical factors $f(I_{1}, I_{3})$. Field equations do not impose any equations on the ``breathing function'' $\lambda$.

\noindent{\bf Proof:} The proof is a matter of direct calculations. We do not quote them. One should substitute a breathing-closed field to (\ref{4.3}) and make use of the fact that the spatial $(n-1)$-tuple $(\phi_{1}, \ldots, \phi_{n-1})$ restricted to integral surfaces of $\phi_{1}\wedge \ldots \wedge \phi_{n-1}$ satisfies $(n-1)$-dimensional equations (\ref{4.3}). It is convenient to start from the simplest model $f=1$. After proving the theorem for this case we can easily generalize it to the non-constant $f$. \rule{5pt}{5pt}

The arbitrariness of the function $\lambda$ reflects the fact that the solutions of (\ref{4.3}) fail to be uniquely determined by their initial data. This lack of determinism is a consequence of the degeneracy of $\mathcal{L}$ and it is rather characteristic for generally-covariant field theories and for all theories invariant under groups involving arbitrary functions. The invariance of $\mathcal{L}$ under the group ${\rm Diff} M$ ``parameterized'' by $n$ arbitrary space-time functions implies that instead of the primary $n^{2}$ degrees of freedom at each spatial point, the system has $n(n-1)$ physical degrees of freedom. The remaining $n$ field functions are gauge variables, whose time dependence is not predicted by dynamical laws.

Let us notice that, in general, the group parameter of $\phi_{0}$ differs from the ``cosmic time'' $T$ measured by the metric tensor $\gamma[\phi]$ along integral curves of $\phi_{0}$ (starting from some fixed maximal integral surface of $\phi_{1}\wedge \ldots \wedge \phi_{n-1}$). The formula (\ref{4.17}) tells us that
$\lambda = A \exp \left(\pm T/\sqrt{n-1}\right)$, where
$A$ is constant. Conversely, if $\tau$ is a parametric time of $\phi_{0}$ and we put $\lambda = A \exp \left(\pm \tau/\sqrt{n-1}\right)$, 
then certainly $T=\tau + {\rm const}$.
Thus, the breathing function $\lambda$ is fixed up to a multiplicative factor by the demand that the one-parameter group generated by $\phi_{0}$ should be identical with the group of time translations.

Let us notice that the metric tensor $\gamma[\phi]$ has rather rich invariance properties. Indeed, it is obvious from the very construction of $\gamma[\phi]$ that there exist at least $n$ independent Killing vectors: $X_{0} = (1/\sqrt{n-1})(\dot{\lambda}/\lambda)\phi_{0}$, $X_{\Sigma} = (1/\lambda) \phi_{\Sigma}=\psi_{\Sigma}$. 
They span a Lie algebra: $\left[X_{\Sigma}, X_{\Lambda} \right] = C^{\Omega}{}_{\Sigma\Lambda}X_{\Omega}$, $\left[X_{0}, X_{\Lambda} \right] =0$. 
Obviously, if we put $\lambda = A \exp \left(\pm \tau/\sqrt{n-1}\right)$, i.e., if the parametric $\phi_{0}$-time coincides with the ``cosmic'' time $T$, then
$X_{0}= \psi_{0}=\phi_{0}$, $X_{\Sigma}= \psi_{\Sigma}=(1/\lambda) \phi_{\Sigma}$.

It is clear that if we fix some origin $e \in M$, then, at least locally, the pseudo-Riemannian manifold $(M, \gamma[\phi])$ can be identified with $\mathbb{R} \times G$, $G$ being the simply-connected Lie group determined uniquely by the structure constants $C^{\Lambda}{}_{\Sigma\Omega}$. Under this identification $e$ becomes the neutral element of the group $\mathbb{R} \times G$, end the metric tensor $\gamma[\phi]$ becomes the direct sum of the natural translationally-invariant metric tensor on $\mathbb{R}$ and of the Killing tensor of $G$. The time-like field $X_{0}$ is transformed into the generator of translations along $\mathbb{R}$, and the vector fields $X_{\Sigma}$ will correspond to the left-invariant vector fields on $G$ (more rigorously, to their natural lifts to $\mathbb{R} \times G$). It is obvious that the right-invariant vector fields on $G$ are also infinitesimal symmetries of the Killing tensor of $G$. These fields are linearly independent of the system of left-invariant fields (linear dependence of fields is here understood in the sense of reals $\mathbb{R}$, i.e., with the use of constant coefficients), because $G$ is semisimple. Therefore, they give rise to additional Killing vectors $X^{*}_{\Sigma}$ of $(M, \gamma[\phi])$ and $\left[X^{*}_{\Sigma}, X^{*}_{\Lambda} \right] = -C^{\Omega}{}_{\Sigma\Lambda}X^{*}_{\Omega}$, $\left[X_{0}, X^{*}_{\Lambda} \right] =0$, $\left[X_{\Lambda}, X^{*}_{\Sigma} \right] =0$.  This means that $(M, \gamma[\phi])$ admits a $(2n-1)$-dimensional Lie group of isometries, isomorphic with the direct product $\mathbb{R} \times G \times G$.

In all calculations it is convenient to use adapted coordinates $\left(\tau, \xi^{\mu}\right)$ or $\left(T, \xi^{\mu}\right)$, where $\xi^{\mu}$ are
constant along integral curves of $\phi_{0}$, and $\tau$ is constant on integral surfaces of $\phi_{1}\wedge \ldots \wedge \phi_{n-1}$. Besides, we assume that, on integral curves of $\phi_{0}$ the quantity $\tau$ coincides with the group parameter of $\phi_{0}$. Obviously, $T$ denotes the ``cosmic time'', 
$dT= \pm \sqrt{n-1} \ d\ln \lambda$. In these coordinates we have $\dot{\lambda}= d \lambda/d \tau$, $\phi_{0}= \partial/\partial \tau$, $\phi^{0}=d \tau$, $X_{0}= \partial/\partial T$, and
\begin{equation}
\gamma= (n-1)\left(\frac{d \ln \lambda}{d \tau}\right)^{2} d \tau \otimes d \tau +\underset{(n-1)}{\gamma}{}_{\mu \nu} d \xi^{\mu} \otimes \xi^{\nu}
= dT \otimes dT + \underset{(n-1)}{\gamma}{}_{\mu \nu} d \xi^{\mu} \otimes \xi^{\nu},
\end{equation}
where $\underset{(n-1)}{\gamma}{}_{\mu \nu}= 4 S^{\lambda}{}_{\mu\kappa}S^{\kappa}{}_{\nu\lambda}= 4 \underset{(n-1)}{S^{\lambda}}{}_{\mu \kappa}\underset{(n-1)}{S^{\kappa}}{}_{\nu \lambda}$ and $\underset{(n-1)}{S^{\lambda}}{}_{\mu \nu}= (1/2) \psi^{\lambda}{}_{\Lambda}(\psi^{\Lambda}{}_{\mu, \nu}-\psi^{\Lambda}{}_{\nu, \mu}) = C^{\Lambda}{}_{\Sigma\Pi} \psi^{\lambda}{}_{\Lambda} \psi^{\Sigma}{}_{\mu}\psi^{\Pi}{}_{\nu}$. Breathing-closed solution are suggestive because they describe the physical $(n-1)$-dimensional space as a micromorphic medium \cite{28}, i.e., continuum of infinitesimal homogeneously deformable grains (concerning the idea of space-time as a micromorphic continuum, cf. \cite{6-1,6-2,6-3,28}).

Let us finish with a few remarks concerning space-time dimension. It turns out that the dimension $4$ is in some sense peculiar. The case $n=1$ would be completely trivial, because $S=0$. If $n=2$, then $S$ need not vanish, but quite independently of $\phi$ the Killing tensor $\gamma[\phi]$ is singular. In three-dimensional manifolds the theory is nontrivial and there exist closed solutions with $S[\phi]$, $\gamma[\phi]$ covariantly constant. These solutions correspond to the Lie algebras $so(3,\mathbb{R}) = su(2)$, $so(1,2;\mathbb{R}) = sl(2,\mathbb{R})$, However, if $(M, \gamma[\phi])$ is to be interpreted as a pseudo-Riemannian space-time structure, then these solutions are inapplicable. Indeed, $su(2)$ is Riemannian (elliptic signature) and $sl(2,\mathbb{R})$ admits closed time-like curves. In three dimensions there are no breathing-closed solutions because two-dimensional Lie algebras are never semisimple. Therefore,
$n=4$ is the lowest dimension compatible with our theory and admitting
Lie-algebraic solutions (breathing-closed fields). This is interesting in itself and perhaps from the point of view of the anthropic principle.

It was mentioned above about the failure of deriving the gradient gauged rule for Jacobi fields, i.e., some small corrections to the $n$-leg field. However, a more careful analysis seems to suggest a hypothesis about some other possibility of dynamical derivation of both the bundle structure and gauge fields from the $n$-leg field in an $n$-dimensional Kaluza-Klein-type Universe. Namely, let us begin from the manifold $M=Y\times G$, where $Y$ is a usual four-dimensional space-time manifold and $G$ is an $(n-4)$-dimensional Lie group. $M$ is to be expected the $n$-dimensional Kaluza-Klein Universe. Let us consider the system of co-frame components given by $e^{A}=e^{A}{}_{\mu}(x)dx^{\mu}$, $e^{\mathbf{R}}=\theta^{\mathbf{R}}{}_{r}(y)dy^{r}+W(y)^{\mathbf{R}}{}_{\mathbf{W}}\phi^{\mathbf{W}}{}_{\mu}(x)dx^{\mu}$, where $\left[W(y)^{\mathbf{R}}{}_{\mathbf{W}}\right]$ denotes the matrix of the co-adjoint representation of $G$, and $e^{A}{}_{\mu}$ is the matrix of the usual four-dimensional gravitational co-tetrad. The matrix $\left[\theta^{\mathbf{R}}{}_{r}\right]$ represents the system of components of the co-adjoint canonical one-form of $G$. The dual contravariant objects $e^{\mu}{}_{A}$, $\theta^{r}{}_{\mathbf{T}}$ are defined by $\theta^{\mathbf{R}}{}_{r}\theta^{r}{}_{\mathbf{S}}=\delta^{\mathbf{R}}
{}_{\mathbf{S}}$ and $e^{A}{}_{\mu}e^{\mu}{}_{B}=\delta^{A}{}_{B}$, therefore, they are explicitly given by
\begin{equation}
e_{A}=e^{\mu}{}_{A}\frac{\partial}{\partial x^{\mu}}-\theta^{r}{}_{\mathbf{R}}(y)U^{\mathbf{R}}{}_{\mathbf{Z}}
\phi^{\mathbf{Z}}{}_{\mu}(x)e^{\mu}{}_{A}(x)\frac{\partial}{\partial y^{r}},\qquad 
e_{\mathbf{R}}=\theta^{r}{}_{\mathbf{R}}(y)\frac{\partial}{\partial y^{r}}.
\end{equation}
The quantities $\theta_{\mathbf{R}}$, $\theta^{\mathbf{R}}$ are related to the basic Lie-algebraic quantities and constants as follows:
$d\theta^{\mathbf{R}}=(1/2)C^{\mathbf{R}}{}_{\mathbf{W}\mathbf{Z}}
\theta^{\mathbf{Z}}\wedge\vartheta^{\mathbf{W}}$, $\left[\theta_{\mathbf{R}},\theta_{\mathbf{Z}}\right]=C^{\mathbf{W}}
{}_{\mathbf{R}\mathbf{Z}}\theta_{\mathbf{W}}$,
where $C^{\mathbf{W}}{}_{\mathbf{R}\mathbf{Z}}$ are structure constants of the gauge groups.
In any case, from the purely formal point of view, it is clear that the field of $n$-legs in $M$ may be exactly equivalent to the space-time foliation and to the gauge fields. But unfortunately, it is not yet clear if any ${\rm GL}(n,\mathbb{R})$-invariant Lagrangian may be responsible for the above fields or for some their reasonable approximation. Nevertheless, it seems quite probable. Then the tetrad $e^{\mu}{}_{A}$ would be responsible for gravitation and $\theta^{r}{}_{\mathbf{R}}$ --- for the gauge fields.
The main idea is that both the (approximate) foliation and the gauge group structures are not fixed a priori, but should be consequences of some dynamical laws of differential equations.

\section{Attempts at gravitational interpretation and spherical solutions}

Above we have given some heuristic arguments for the hypothesis that the generally covariant and affinely invariant Lagrangian dynamics of the ``tetrad'' field could be useful as a geometric model of some fundamental interaction. Then the general mathematical formalism was given, and finally we have
presented two simplest solutions (closed and breathing-closed fields). Thus, certainly, our field equations are non-empty. However, this is not yet physics. 

A priori the following possibilities of physical interpretations seem to be possible and should be investigated:
1.\ Modified gravitation theory using as a carrier of interaction the quadruple of vector particles instead of the graviton (tensor particle). 
2.\ The theory of electroweak interactions. This conjecture is motivated by the fact that the fundamental object in our formalism is a quadruple
of vector bosons and that we deal with 12 degrees of freedom, just as in the boson sector of the standard Salem-Weinberg model. It is
also interesting that in four dimensions our breathing-closed solutions are given by the $U(2)$-Lie algebra, thus, the dynamics of small
corrections to these solutions should somehow ``feel'' certain geometry based on $U(2)$.
3.\ Perhaps our formalism could unify gravitation and electroweak interactions as carried by the same agent, i.e., the quadruple of vector bosons.
Obviously, it is also possible that there is no physical counterpart of our concepts or that such a counterpart does exist but not among physical objects known today. 

We finish this paper with some remarks concerning the possibility of interpretation of our field equations (\ref{4.3}) in gravitational
terms. Let us observe that there are some reasons towards this. Namely, let us put $g[e]_{ij}=4AS^{a}{}_{bi}S^{b}{}_{aj}$ ($A$ being a constant) as the metric tensor of the physical space-time. And now let $R_{ij}$ denote its Ricci tensor, and $R=g^{ij}R_{ij}$ --- its scalar curvature. Let us also assume that the field of frames $e$ is a basis of some semisimple Lie group. One can show \cite{Hal_84} that
\begin{equation}
R_{ij}-\frac{1}{2}Rg_{ij}=\frac{2-n}{8A}g_{ij}.
\end{equation}
But those are the Einstein equations with the cosmological constant given by $\Lambda=-1/(4A)$ when $n=4$. This means that in some region of solutions close to a Lie group, there exists some correspondence with Einstein theory (although with the cosmological constant). Other arguments were shown in the papers of P. Godlewski \cite{PG_02,PG_03_1,PG_05,PG_08,PG_10,PG_13}.

Let us try to interpret the quantity $\gamma_{ab}$ or $g_{ab}=\gamma_{ab}+\mu\gamma_{a}\gamma_{b}$ as the genuine metric tensor occurring, e.g., in equations of electrodynamics. We shall assume that the interaction of macroscopic matter with this metric tensor is described by the Einstein scheme. In particular, in the Newton limit, we interpret $\gamma_{00}$ as $1+2\varphi$, where the $3$-dimensional scalar $\varphi$ is the usual gravitational potential.
Obviously, in breathing-closed solutions the Newtonian potential is trivial, because $\gamma_{00}$ is constant. The only gravitational
effect predicted by such solutions is connected with the $3$-dimensional spatial curvature, which manifests itself through deviation of world-lines
of particles with internal degrees of freedom from the geodetic shape \cite{27,JJS+BG_10}. Thus, although the space-time $\left(M,\gamma\right)$
is curved, it is as ``flat'' as possible on the basis of equations (\ref{4.3}) in four dimensions. To be acceptable as an alternative
gravitation theory, our equations (\ref{4.3}) must possess solutions with nontrivial $\gamma_{00}$-components and with a reasonable Newtonian
asymptotic. In particular, it is necessary that there exist solutions $\phi$ whose Killing tensors $\gamma\left[\phi\right]$ are spherically-symmetric
and qualitatively similar to the well-established Schwarzschild solution of Einstein equations. Thus, from now on, we focus our attention on spherically symmetric fields.

Let us put $M=\mathbb{R}^{4}=\mathbb{R}\times\mathbb{R}^{3}$ and denote the natural coordinates by $x^{i},\: i=0,1,2,3$. The coordinate
$x_{0}$, denoted also by $t$, is to be a ``time-like'' variable, whereas $x^{\mu}$, $\mu=1,2,3$, will be ``spatial'' coordinates (in
the sequel the Greek indices always run over the range $1,2,3,$ whilst the Latin ones run over the range $0,1,2,3$). This means that we restrict
ourselves to such tetrad field $\phi$ that the vector fields $\partial/\partial x^{0}$, $\partial/\partial x^{\mu}$
are respectively time-like and space-like with respect to the Killing metric tensor$\gamma\left[\phi\right]$. We shall also use the spherical
coordinates $r,\theta,\varphi$ in $\mathbb{R}^{3}$ and the versor components $n^{\mu}=x^{\mu}/r$.
Isotropic tetrad fields will be sought in the following form:
1) the ``temporal'' leg: 
\begin{equation}
\phi_{0}=K\left(r,t\right)\frac{\partial}{\partial t}+J\left(r,t\right)x^{\mu}\frac{\partial}{\partial x^{\mu}}=K\left(r,t\right)\frac{\partial}{\partial t}+J\left(r,t\right)r\frac{\partial}{\partial r},\label{5.1a}
\end{equation}
2) the ``spatial'' legs, $\Lambda=1,2,3$: 
\begin{eqnarray}
\phi_{0} & = & I\left(r,t\right)x_{\Lambda}\frac{\partial}{\partial t}+\left[F(r,t)\delta^{\mu}{}_{\Lambda}+G(r,t)x^{\mu}x_{\Lambda}+ H(r,t)\varepsilon_{\Lambda\nu}{}^{\mu}x^{\nu}\right]\frac{\partial}{\partial x^{\mu}}\nonumber\\
&=&rI\left(r,t\right)n_{\Lambda}\frac{\partial}{\partial t} +  \left(F(r,t)+r^{2}G(r,t)\right)n_{\Lambda}\frac{\partial}{\partial r} \nonumber\\
&-&  \frac{1}{r}R(r,t)\varepsilon_{\Lambda\mu}{}^{\nu}n^{\mu}D_{\nu}+H(r,t)D_{\Lambda},\quad\label{5.1b-c}
\end{eqnarray}
where $D_{\lambda}=\varepsilon_{\lambda\alpha}{}^{\beta}x^{\alpha}\partial/\partial x^{\beta}$ and $F$, $G$, $H$, $I$, $J$, $K$ are certain shape functions depending only on the variables $(t,r)$. The raising and lowering of indices at $\delta,\, x$, and $\varepsilon$, is understood in these formulas in the trivial $\mathbb{R}^{3}$-Kronecker sense; it is used only for ``cosmetic'' purposes, e.g., to avoid ``graphical'' conflicts with the summation
convention.

The formulas above describe the most general tetrad field co-variant with respect to the group ${\rm SO}(3,\mathbb{R})$ acting as a natural
diffeomorphism group of $M=\mathbb{R}\times\mathbb{R}^{3}$. The term ``covariant'' is understood in the sense that the components $\phi^{i}{}_{A}$
satisfy the conditions: 
$\phi^{\mu}{}_{\Lambda}(t,Rx) = R^{\mu}{}_{\nu}\phi^{\nu}{}_{\Sigma}(t,x)R^{-1\Sigma}{}_{\Lambda}$, $\phi^{0}{}_{\Lambda}(t,Rx)  =  \phi^{0}{}_{\Sigma}(t,x)R^{-1\Sigma}{}_{\Lambda}$,
$\phi^{\mu}{}_{0}(t,Rx)=R^{\mu}{}_{\nu}\phi^{\nu}{}_{0}(t,x)$,
$\phi^{0}{}_{0}(t,Rx) =  \phi^{0}{}_{0}(t,x)$
for any $\mathbb{R}\in {\rm SO}(3,\mathbb{R})$ and $t\in\mathbb{R},\, x\in\mathbb{R}^{3}$.
The Killing metric tensor is then also spherically symmetric, 
$\gamma_{00}=\gamma_{00}(t, r)$,
$\gamma_{0\mu}=\gamma_{\mu 0}=\gamma_{0}\left(t,r\right)x_{\mu}=
r\gamma_{0}\left(t,r\right)n_{\mu}$, and
$\gamma_{\mu\nu}=\gamma_{\nu\mu}=
\gamma[0](t,r)\delta_{\mu\nu}+\gamma[2](t,r)x_{\mu}x_{\nu}=
\gamma[0](t,r)\delta_{\mu\nu}+r^{2}\gamma[2](t,r)n_{\mu}n_{\nu}$,
where $\gamma_{00},\,\gamma_{0},\gamma\left[0\right],\gamma\left[2\right]$
are certain functions of $(t,r)$, built in a rational way of the
above functions $F$, $G$, $H$, $I$, $J$, $K$ and their first-order derivatives.
In other words, 
$\gamma_{00}(t,Rx) =  \gamma_{00}(t,x)$,
$\gamma_{0\mu}(t,Rx) =  \gamma_{0\nu}(t,x)\left.R^{-1}\right.^{\nu}{}_{\mu}$,
$\gamma_{\mu\nu}(t,Rx)  =  \gamma_{\alpha\beta}(t,x)\left.R^{-1}\right.^{\alpha}{}_{\mu}
\left.R^{-1}\right.^{\beta}{}_{\nu}$.
Substituting the above form of $\phi$ to our field equations (\ref{4.3})
we obtain a system of $6$ partial differential equations for 6 functions
$F$, $G$, $H$, $I$, $J$, $K$ of two variables $(t,r)$. Indeed, equations (\ref{4.3}) have the form $K_{i}{}^{j}=0$, where $K$ is the mixed tensor density
of weight one given by 
$K_{i}{}^{j}=\nabla_{k}H_{i}{}^{ji}+2S^{m}{}_{mk}H_{i}{}^{jk}$.
The isotropic structure of $\phi$ implies that $K$ is also isotropic,
$K_{0}{}^{0}=K_{0}{}^{0}\left(t,r\right)$,
$K_{0}{}^{\mu}=K_{0}\left(t,r\right)x^{\mu}=rK_{0}\left(t,r\right)n^{\mu}$,
$K_{\mu}{}^{0}=K^{0}\left(t,r\right)x_{\mu}=rK_{0}\left(t,r\right)n_{\mu}$,
$K_{\mu}{}^{\nu}=K[0](t,r)\delta_{\mu}{}^{\nu}+K[1](t,r)
\varepsilon_{\mu}{}^{\nu k}x_{k}+K[2](t,r)x_{\mu}x^{\nu}= K[0]\delta_{\mu}{}^{\nu}$ $+rK[1]\varepsilon_{\mu}{}^{\nu k}n_{k}+r^{2}K[2]n_{\mu}n^{\nu}$, where the shape functions $K_{0}{}^{0}$, $K_{0}$, $K^{0}$, $K[0]$, $K[1]$, $K[2]$ depend only on variables $(t,r)$.
In this way, field equations (\ref{4.3}) applied to spherically symmetric
tetrads $\phi$ reduce to the following system of equations: 
\begin{equation}
K_{0}{}^{0}=0,\quad K_{0}=0,\quad K^{0}=0,\quad K[0]=0,\quad K[1]=0,\quad K[2]=0.\label{5.7}
\end{equation}

Let us notice that coordinates $x^{i}$ are not uniquely fixed by the demand that $\phi$ should have the shape (\ref{5.1a}), (\ref{5.1b-c}).
Indeed, any smooth change of coordinates on the $(t,r)$-plane is admissible, i.e.,
\begin{equation}
\left(t,r\right)\rightarrow\left(\overline{t},\overline{r}\right)=
\left(a\left(t,r\right),b\left(t,r\right)\right).\label{5.8}
\end{equation}
Such transformations do not affect either the form (\ref{5.1a}), (\ref{5.1b-c}) or the field equations (\ref{4.3}), (\ref{5.7}); this is a consequence
of general covariance. Transformation formula (\ref{5.8}) involves two arbitrary functions $a,\, b$ of two variables $(r,t)$. Thus,
the system of shape functions $(F,\ldots,K)$ is redundant, because in principle two of them can be given any a priori prescribed form.

In Einstein theory field equations together with a proper choice of coordinates in the $(t,r)$-plane enable us to eliminate the time
variable; this elimination reduces the equations for spherically symmetric fields to ordinary differential equations for the shape
functions. This implies in particular that the Schwarzschild solution (gravitational field of a point mass) is static. In our affinely-invariant
theory of the tetrad field it does not seem possible to get rid of the time variable by a change of coordinates (\ref{5.8}). Nevertheless,
on the basis of analogy with breathing-closed solutions we can show that there exists a natural class of isotropic solutions described
by ordinary differential equations. First of all, let us observe that breathing-closed fields $\phi$ can be alternatively
described by the formulas: $\phi_{0}=\lambda\left(t\right)\psi_{0}$, $\phi_{\Lambda}=\lambda\left(t\right)\psi_{\Lambda}$.
This modification does not influence the formulas for $\gamma$, in particular, the parameter $t$ coincides with the physical time measured
along orbits of $\phi_{0}$ iff $\lambda=\mathrm{const}\cdot\exp(+t/\sqrt{3})$.
Thus, it is natural to postulate the following form of isotropic solutions:
\begin{equation}
\left(F,G,H,I,J,K\right)\left(r,t\right)=e^{\alpha t}\left(f(r),g(r),h(r),i(r),j(r),k(r)\right),\label{5.10}
\end{equation}
$\alpha$ being constant. In fact, it turns out that this conjecture is correct. If we substitute (\ref{5.10}) into (\ref{5.7}), then
the time variable drops out of equations and we obtain a system of six second-order ordinary differential equations for six functions
$(f,\ldots,k)$ of the radial variable. This system involves $\alpha$ as a parameter, however, no restrictions for the value of $\alpha$
are imposed by equations. Ordinary differential equations satisfied by $(f,\ldots,k)$ are strongly
nonlinear. Their left-hand sides are rather complicated rational functions of the shape functions and of their first-order derivatives; obviously,
the second derivatives occur in a linear way.

It is interesting that the overall exponential time dependence of the tetrad implies that coefficients $\Gamma^{i}{}_{jk}$ of the
parallelism connection do not depend on time. Therefore, the torsion coefficients $S^{i}{}_{jk}$ and all their algebraic functions, e.g.,
components of the Killing tensor $\gamma_{ij}$ are time-independent. This means that the pseudo-Riemannian manifold $\left(\mathbb{R}\times\mathbb{R}^{3},\gamma[\phi]\right)$ is stationary (but in general, non-static; $\gamma_{0\mu}\neq 0$). 
Conditions (\ref{5.10}) impose certain additional restrictions on the coordinate system, nevertheless there exists still some rather
large gauge freedom. Namely, the exponential factorization (\ref{5.10}) is preserved by the following deformations of coordinates:
1) radial variable deformations, 
\begin{equation}
\left(t,r\right)\rightarrow\left(\overline{t},\overline{r}\right)=
\left(t,\omega(r)\right),\label{5.11a}
\end{equation}
2) $r$-dependent time translations, 
\begin{equation}
\left(t,r\right)\rightarrow\left(\overline{t},\overline{r}\right)=
\left(t+\varepsilon(r),r\right),\label{5.11b}
\end{equation}
where $\omega$ and $\varepsilon$ are in principle arbitrary functions of $r$.
These transformations preserve the exponentially-factorised shape of $\phi$ and result in the following transformations
of $(f,\ldots,k)$:
\begin{eqnarray}
1)&& \overline{f}(\omega(r)) =  f(r)\frac{\omega(r)}{r},
\quad \overline{h}(\omega(r))  =  h(r),\quad \overline{j}(\omega(r))  =  j(r)r\frac{d\ln\omega(r)}{dr},\nonumber \\
&&\overline{g}(\omega(r)) =  g(r)\frac{r^{2}}{\omega(r)}\frac{d\ln\omega(r)}{dr}
+\frac{f(r)}{\omega(r)}\left(\frac{d\ln\omega(r)}{dr}
-\frac{1}{r}\right),\nonumber\\
&&\overline{i}(\omega(r))  =  i(r)\frac{r}{\omega(r)},\quad
\overline{k}(\omega(r)) =  k(r).\label{5.12a}\\
2)&& \overline{f}=\exp\left(-\alpha\varepsilon\right)f,\
\overline{g}=\exp\left(-\alpha\varepsilon\right)fg,\
\overline{h}=\exp\left(-\alpha\varepsilon\right)h,\
\overline{j}=\exp\left(-\alpha\varepsilon\right)j,\nonumber \\
 && \overline{i}=\exp\left(-\alpha\varepsilon\right)\left(i+\frac{1}{r}
\left(f+gr^{2}\right)\frac{d\varepsilon}{dr}\right),\
\overline{k}=\exp\left(-\alpha\varepsilon\right)\left(k+rj
\frac{d\varepsilon}{dr}\right).\label{5.12b}
\end{eqnarray}
These transformation rules involve two arbitrary functions, thus, it is in principle possible to deform two of the six functions, $(f,\ldots,k)$
to any a priori fixed shape. Therefore, when some gauge is fixed, we are dealing with a system of six second-order ordinary differential equations
imposed on four shape functions. As mentioned above, these equations are extremely complicated and when written down explicitly are completely
dark. It is rather hard to expect rigorous solutions in analytical form. We suppose that more realistic and physically more interesting
is the following problem: to estimate the asymptotic behaviour of the shapes functions $(f,\ldots,k)$ and the Killing tensor $\gamma[\phi]$
about the origin $r=0$. This is necessary if we intend to compare our model with well-established consequences of Einstein theory of
gravitation and with Newton theory.

The strong nonlinearity of our field equations and the ``Born-Infeld structure'' of $L$ enable us to conjecture that perhaps
there exist solutions finite at $r=0$. It is interesting whether there exist black holes in our model, in particular, whether there
exists horizon-effect with hypothetical solutions finite at $r=0$. As yet, we are unable to answer such questions.

It is much more easy to discuss the correspondence with Newton potential and with Schwarzschild metric in the weak-field approximation. We
have seen in the above sections that there exist certain explicitly known isotropic solutions, namely, breathing-closed solutions corresponding to $su(2)$-Lie
algebra. The manifold $M$ becomes then locally identical with $\mathbb{R}\times SU(2)=\mathbb{R}\times S^{3}$,
or with $\mathbb{R}\times {\rm SO}(3,\mathbb{R})$. Let us parametrize ${\rm SO}(3,\mathbb{R})$
with the help of the rotation vector $\overline{\rho}=\left(\rho_{1},\rho_{2},\rho_{3}\right)$
(canonical coordinates of the first kind on ${\rm SO}(3,\mathbb{R})$. This parametrization identifies ${\rm SO}(3,\mathbb{R})$ with the closed sphere
$\rho\leq\pi$ in $R^{3}$ (where, obviously, $\rho=\sqrt{\left.\rho_{1}\right.^{2}+\left.\rho_{2}\right.^{2}+
\left.\rho_{3}\right.^{2}}$) with the proviso that antipodal points on the surface $\rho=\pi$ are identified. Similarly, $SU(2)$ becomes the sphere $\rho\leq 2\pi$ with the proviso that the whole surface $\rho=\pi$ is identified with $-I$, $I$ being $2\times 2$ identity matrix. The shape functions
corresponding to the ${\rm SO}(3,\mathbb{R})$-breathing-closed solutions are given by
$f_{0}=\left(\rho/2\right)\mathrm{ctg}\left(\rho/2\right)$, $g_{0}=\left(1/\rho^{2}\right)\left(1-\left(\rho/2\right)\mathrm{ctg}
\left(\rho/2\right)\right)$, $h_{0}=\pm 1/2$, $i_{0}=0$, $j_{0}=0$, $k_{0}=1$.
This parametrization is inconvenient because it leads to expressions in which trigonometric functions occur simultaneously with algebraic
ones. Thus, it is better to use the vector of finite rotation as a parametrization of ${\rm SO}(3,\mathbb{R})$; it is related to canonical
coordinates through the formula  $\overline{r}/r=\overline{\rho}/\rho$, $r=\mathrm{tg}\left(\rho/2\right)$.
This parametrization identifies ${\rm SO}(3,\mathbb{R})$ with the projective space $P\mathbb{R}^{3}$, rotations by $\pi/2$ being represented
by points at infinity. Trigonometric functions are eliminated and the ${\rm SO}(3,\mathbb{R})$-breathing-closed solution is given by the
following very simple shape functions: 
$f_{0}=g_{0}=h_{0}=1/2$, $i_{0}=j_{0}=0$, $k_{0}=1$.
The corresponding Killing tensor has the form: 
$ds^{2}=3\alpha^{2}dt^{2}-[8/(1+r^{2})^{2}]dr^{2}-[8r^{2}/(1+r^{2})](d\vartheta^{2}+\sin^{2}\vartheta d\varphi^{2})$.

Let us now consider small spherically-symmetric perturbations of the above breathing-closed solutions. In other words, the shape functions
are put in the following form:
\begin{equation}
f=\frac{1}{2}+\varphi\quad g=\frac{1}{2}+\gamma,\quad 
h=\frac{1}{2}+\chi,\quad i=\mu,\quad j=\nu,\quad k=1+\kappa,\label{5.17}
\end{equation}
where $\varphi,\,\gamma,\,\chi,\,\mu,\,\nu,\,\kappa$ are small corrections depending only on the variable $r$. Substituting (\ref{5.17}) to
(\ref{5.7}) and neglecting higher-order terms we obtain a system of six linear ordinary differential equations imposed on six functions
$\left(\varphi,\ldots,\kappa\right)$. The general covariance enables us to eliminate two of these functions.

Let us consider an infinitesimal transformation (\ref{5.11a}), (\ref{5.11b}), i.e., we put $\omega=1+\eta$, and assume that $\eta$ and $\varepsilon$
are small. Linearizing expressions (\ref{5.12a}), (\ref{5.12b}) with respect to $\eta,\,\varepsilon,\,\varphi,\ldots,\kappa$, we obtain
the following transformation rules for infinitesimal shape functions:
\begin{eqnarray}
1)\qquad && \overline{\varphi}=\varphi+\frac{1}{2r}\eta,\quad\overline{\gamma}=\gamma-
\frac{1+2r^{2}}{2r^{3}}\eta+\frac{1+r^{2}}{2r^{2}}\frac{d\eta}{dr},\nonumber\\
 && \overline{\chi}=\chi,\quad\overline{\mu}=\mu,\quad\overline{\nu}=\nu,
\quad\overline{\kappa}=\kappa,\label{5.18a}\\
2)\qquad && \overline{\varphi}=\varphi-\frac{1}{2}\alpha\varepsilon,\quad\overline{\gamma}=
\gamma-\frac{1}{2}\alpha\varepsilon,\quad\overline{\chi}=\chi-\frac{1}{2}
\alpha\varepsilon,\nonumber\\
 && \overline{\mu}=\mu+\frac{1}{2r}\left(1+r^{2}\right)\frac{d\varepsilon}{dr},
\quad\overline{\nu}=\nu,\quad\overline{\kappa}=\kappa-\alpha\varepsilon.\label{5.18b}
\end{eqnarray}
It is interesting that among all infinitesimal shape functions $\left(\varphi,\ldots,\kappa\right)$, $\nu$ is the only purely physical quantity invariant under coordinate gauge transformations (\ref{5.18a}), (\ref{5.18b}). The most convenient gauge is $\mu=0,\,\gamma=0$, because second derivatives of these functions do not enter linear equations for small corrections. One can show that in linear approximation we have
\begin{equation}
\gamma_{00}=3\alpha^{2}+\frac{2\alpha\left(r^{2}-3\right)}{1+r^{2}}\nu-2\alpha r\frac{d\nu}{dr}.\label{5.19}
\end{equation}
It is seen that in this approximation the gravitational potential $\gamma_{00}$ is controlled by the shape function $\nu$ alone, i.e.,
by the ``spatial'' components of $\phi_{0}$, just by the only gauge-independent shape function. This is correct, if $\nu$ is to represent gravitational
scalar potential (scalar in the $3$-dimensional sense, of course).

It has been mentioned above that it is difficult to decide a priori whether the macroscopic metric tensor should be identified with $\gamma_{ij}$
or with a more general expression $g_{ij}=\lambda\gamma_{ij}+\mu\gamma_{i}\gamma_{j}$ (e.g., $A\gamma_{ij}+B\gamma_{i}\gamma_{j}$, $A,\, B$ being the constants occurring in our ``Born-Infeld'' Lagrangian). Our weak-field gravitational test presented above is neutral with respect to this
question, because in a linear approximation $g_{00}$ is proportional to $\gamma_{00}$: $g_{00}=\left(\lambda+3\mu\right)\gamma_{00}$.

Calculations leading to linear equations for small corrections $\left(\varphi,\ldots,\kappa\right)$ are very strenuous. The final equations are also rather complicated, thus, we do not quote them here. They will be reported in another paper. Obviously, the explicit form of equations depends on the choice
of Lagrangian $L$. Our calculations were based on the Born-Infeld model, $L=\sqrt{t}$, where $t_{ij}=A\gamma_{ij}+B\gamma_{i}\gamma_{j}+C\Gamma_{ij}$. Calculations accompanying Lagrangian $L=f\sqrt{t}$ with non-constant
dynamical factors $f$ are so complicated that, in our opinion, they are practically impossible to be carried out ``on foot'', without
using computer-programmed formal processes.

Our linear equations have non-constant coefficients, thus, to obtain any explicit result we have to use the Frobenius power-series method.
We are especially interested in the asymptotic behaviour of $\gamma_{00}$, (\ref{5.19}) about the origin $r=0$, thus, all shape functions $\left(\varphi,\ldots,\kappa\right)$ are represented as power series of the variable $r$. Luckily, $r=0$ is a proper singular point of our system of equations. The asymptotics
of solutions at the origin is determined by the characteristic equation of the system. Unfortunately, even on the linearization level, we were not yet successful
in answering the questions formulated above. Everything we can do now is to report our difficulties and hypotheses.

For generic values of the parameters $A,\, B,\, C,$ our characteristic equation has two solutions: $p=0,\, p=-3$. One can easily show that
the most general solution corresponding to the exponent $p=0$ has the form: 
\begin{equation}
\varphi=2c,\quad\gamma=0,\quad\chi=c,\quad\kappa=d,\quad\mu=0,\quad
\nu=0\label{5.20}
\end{equation}
modulo the gauge (\ref{5.18a}), (\ref{5.18b}); the quantities $c,\, d$ are arbitrary constants. For generic values of these parameters the
solution (\ref{5.20}) cannot be reduced to the trivial one by any gauge transformation (\ref{5.18a}), (\ref{5.18b}). The exceptional,
essentially trivial solutions correspond to the choice $d=2$c. When parameters $c,\, d$ are small, then any solution (\ref{5.20}) also is uniformly small in $M$, in agreement with our very assumption that $\left(\varphi,\ldots,\nu\right)$ are small corrections to breathing-closed $SU(2)$-solutions. Therefore, linear approximation seems to confirm our hypothesis concerning spherically-symmetric solutions finite at the origin $r=0$ (Born-Infeld finiteness effect). This is interesting in itself, but, unfortunately, such solutions do not predict the scalar gravitational potential, because $\nu=0$ and consequently $\gamma_{00}=\mathrm{const}$ (cf.\ (\ref{5.19})). As mentioned in previous remarks, we can try to interpret the quantity 
\begin{equation}
H^{ij}=\left|\gamma\right|^{-1}H_{a}{}^{ib}H_{b}{}^{ja}\label{5.21}
\end{equation}
as another candidate for the contravariant metric tensor. However, this does not seem to help us with the difficulty of the vanishing
gravitational potential.

As yet we have no sure results concerning the second characteristic exponent, $p=3$. To obtain them one has to perform very complicated
calculations. Solutions corresponding to $p=-3$ should be sought as power series modified by logarithmic terms constructed with the
help of (\ref{5.20}), because characteristic exponents differ by an integer. At first sight the exponent $p=-3$ seems encouraging,
because characteristic equation does not impose then any restrictions $\chi_{0},\,\nu_{0}$ (on the contrary, $\varphi_{0}=0,\,\kappa_{0}=0$)
and equation (\ref{5.19}) implies that the power series for $\gamma_{00}$
starts from the term $\left(a+b/r\right),\, a,\, b$ being constants. This is just the typical Newton-Schwarzschild behaviour. However,
things are more complicated. Namely, the gauge freedom (\ref{5.18a}), (\ref{5.18b}) 
implies that iteration equations are overdetermined: in the $n$-th step of iteration we have six linear equations connecting four unknowns
$\varphi_{n},\,\chi_{n},\,\kappa_{n},\,\nu_{n}$ with earlier coefficients $\varphi_{m},\,\chi_{m},\,\kappa_{m},\,\nu_{m},\, m<n$ (and all coefficients
occurring on a given iteration step have the same parity, i.e., $m,\, n$ are either both even or both odd). Therefore, a priori, it is rather
natural to expect that the overdetermined system for $\varphi_{n},\,\chi_{n},\,\kappa_{n},\,\nu_{n}$
will be contradictory unless the earlier coefficient $\varphi_{m},\,\chi_{m},\,\kappa_{m},\,\nu_{m},\, m<n$,
satisfy certain additional linear homogeneous equations. This means that we obtain certain linear conditions for coefficients $\chi_{0},\,\nu_{0}$,
and it may happen quite easily that $\chi_{0}=\nu_{0}=0$, i.e., odd-powers
solutions of (\ref{5.7}) are trivial. It it happens so, then certainly there is no correspondence with Newton potential and with Schwarzschild
solution. Even if there exist nontrivial solutions composed of even powers of $r$ and of the aforementioned logarithmic terms, they cannot
help this failure (they can merely make $\gamma_{00}$ nontrivial, but certainly non-Newtonian). There is certainly no trivialization
in the first even-order step following the characteristic equation, $n=2$, because quite independently of the choice of constants $A,\, B,\, C,$
two of the six equations for $\varphi_{2},\,\chi_{2},\,\kappa_{2},\,\nu_{2}$
are algebraic consequences of the remaining four ones, thus there is no restriction for $\chi_{0}$ and $\nu_{0}$. However, the complicated
structure of equations corresponding to the exponent $p=-3$ prevented us understanding what happens at higher iteration steps.

Let us mention that one should be careful when interpreting solutions corresponding to $p=-3$, because they are singular at $r=0$, in
contradiction with our primary assumption that they are ``small''. Thus, at the present stage of our work, the problem of the Newton-Schwarz\-schild
limit seems to be open and far from being solved. But what will it mean, if it happens however that for $p=-3$ there are only trivial
solutions for $(\varphi,\,\chi,\,\kappa,\,\nu)$? A priori there are the following possibilities of interpreting such a negative answer:
1. ${\rm GL}(4,\mathbb{R})$-invariant models are essentially useless for describing macroscopic gravitational phenomena. 
2. Perhaps there exist solutions of equations (\ref{5.7}) with $\nu\neq 0$ impossible to be found with the help of the Frobenius method. 
3. The coupling of the field $\phi$ with matter should be non-Einsteinian (in the sense of the metric field $\gamma\left[\phi\right]$), e.g., we should try to interpret the quantity $H_{ij}$ (\ref{5.21}) as a macroscopic metric tensor interacting with matter. 
4. Perhaps some modified Lagrangians, e.g., $\left(aI_{1}+bI_{3}+c\right)\sqrt{\left|t\right|}$ admit solutions with non-vanishing (in particular, Newton-like) $\nu$.
5. Perhaps there are solutions of (\ref{5.7}) with non-constant (in particular, Schwarzschild-like) $\gamma_{00}$, but they cannot be
found on the basis of linearized equations. Indeed, it is known that the linearization procedure is non-reliable in generally-covariant
theories with background solutions admitting infinitesimal symmetries. The non-constancy of $\gamma_{00}$ would he an essentially nonlinear
effect. We would be inclined to believe in the last hypothesis. In any case, it is clear that all questions concerning black-holes and horizons in pseudo-Riemannian manifold $\left(M,\gamma[\phi]\right)$ can be properly answered only on the basis of a non-linear analysis of (\ref{5.7}). 

Let us also mention interesting and successful efforts to find and discuss general isotropic solutions in other alternative models of gravity,
e.g., in metric-teleparallel models and in quadratic metric-affine theories \cite{1,10}. Both the Newtonian limit and the correspondence
with Schwarzschild space-time were shown to exist there, and in addition certain ideas concerning microphysical aspects of those solutions
(confinement problem) were formulated. Unfortunately, it does not seem possible to follow methods used in mentioned papers, because
nonlinearity of our field equations is much stronger.

\section*{Final remarks}

Certain objections may be raised against the general form of solutions discussed here, even if they happen
to be satisfactory from the gravitational point of view. Namely, there is something unpleasant in their real-exponential time dependence.
The fields $\phi_{A}$ corresponding to such solutions become infinite when $t\rightarrow+\infty$ or $t\rightarrow-\infty$, depending on
the sign of $\alpha$. Obviously, the metric tensor $\gamma[\phi]$, just as any other ${\rm GL}(4,\mathbb{R})$-invariant tensorial quantity
built of $\phi$, ``does not feel'' this infinity, moreover, it is independent of time. In some sense the exponential breathing is a
non-physical breathing of gauge variables. Nevertheless, there is something aesthetically unsatisfactory in this kind of time-dependence.
It would seem much more natural if the breathing was oscillatory, i.e., if $\alpha$ was purely imaginary. Obviously, to have such solutions,
we must admit from the very beginning complex vector fields $\phi_{A}$, complex connections, and complex torsion. We suppose that Lagrangians
of the form $\sqrt{\left|t\right|}$ with $t$ bilinear in $S$ and $S^{*}$ could admit ``breathing'' solutions periodic in time (cf.\ \cite{PG_02,PG_03_1,PG_05,PG_08,PG_10,PG_13}).

\end{document}